\documentclass[10.5pt]{article}

\usepackage[T1]{fontenc}
\usepackage{lmodern}

\usepackage{graphicx}
\graphicspath{{./figs/}}  
\usepackage[usenames]{color}
\usepackage{amsmath,amssymb,amsfonts,amsthm,amscd}	
\usepackage{bm,braket,array}
\usepackage{multirow}
\usepackage[all]{xy}
\usepackage{slashed}
\usepackage[format=hang,justification=raggedright]{caption} 

\theoremstyle{definition}

\theoremstyle{remark}

\def\re{{\rm Re\,}}
\def\im{{\rm Im\,}}
\def\pf{{\rm Pf\,}}
\def\mod{{\rm\ mod\ }}

\def\p{\partial}

\def\ker{{\rm Ker\,}}
\def\coker{{\rm Coker\,}}

\newcommand{\s}{\sigma}
\newcommand{\g}{\gamma}
\newcommand{\G}{\Gamma}

\newcommand{\C}{\mathbb{C}}

\newcommand{\Z}{\mathbb{Z}}

\newcommand{\bk}{\bm{k}}

\def\widebar{\accentset{{\cc@style\underline{\mskip10mu}}}} 
\def\wideubar{\underaccent{{\cc@style\underline{\mskip10mu}}}} 

\title{Variants of the symmetry-based indicator}
\author{
Ken Shiozaki\thanks{ken.shiozaki@yukawa.kyoto-u.ac.jp}\\
\ \\
{\it Yukawa Institute for Theoretical Physics, Kyoto University, Kyoto 606-8502, Japan}
}
\date{\today} 
\begin{document} 
\maketitle
\begin{abstract}
The symmetry-based indicator [H. C. Po, A. Vishwanath, H. Watanabe, Nat. Commun. 8, 50 (2017)] is a practical tool to diagnose topological materials in the band theory. 
In this note, we present two directions to generalize the symmetry-based indicator for other classes of topological materials. 
The one is for superconductors. 
The careful definition of the atomic insulators and the trivial vacuum Hamiltonian yields the symmetry-based indicators specific to superconductors. 
The other is for ingap boundary states. 
The quotient of the group of atomic insulators by a subset of atomic insulators such as those localized at the interior of the unit cell gives us the symmetry-based indicator for detecting ingap corner, hinge, and surface states. 
\end{abstract}

\setcounter{tocdepth}{2}

\parskip=\baselineskip

\section{Introduction}
The concept of the symmetry-based indicator (SI)~\cite{Po230, Haruki1651} is summarized as something nontrivial that is detectable from irreps at high-symmetric points. 
The SI is a realistic tool to diagnose a nontrivial band topology in several points listed below. 
First, the SI is easy to compute for a given band structure. 
The SI is composed only of the data of the numbers of irreps at high-symmetric points, which is the generalization of early works for the Fu-Kane-type formulas.~\cite{FuKaneInversion, FuBerg, SatoOddParity, ChenRotation}
The second issue is that in the presence of magnetic space group symmetry it seems impractical to list all topological invariants in the Brillouin zone (BZ). 
The computation of the $K$-group via the Aityah-Hirzebruch spectral sequence showed there exist a lot of topological invariants defined on sub-skeletons in the BZ.~\cite{KS_Atiyah}
The explicit construction of topological invariants is a case-by-case problem for each magnetic space group. 
See Refs.~\cite{ChenGlide, KSGlide, KSGlide2, KS_Atiyah} for some topological invariants beyond the ten-fold classification.~\cite{RyuTenFold, KitaevPeriodic}
The third point is the mismatch in between the topology in the momentum space and that in the real space.~\cite{Po230, TQC}
In the real space, the atomic insulators, which are just occupation states of atomic orbitals, are less nontrivial in the viewpoint of topology, however, in the momentum space, inequivalent atomic insulators are sometimes distinguished by a complicated topological invariant. 
The SI is designed not to detect atomic insulators. 

After the proposal and the enumeration of the SI for 1651 magnetic space groups in spinless and spinful electric systems,~\cite{Po230, Haruki1651} the explicit expressions of the SIs for for 230 space groups and the implications of nontrivial values of the SI ware explored,~\cite{SongAII, SongAI, KhalafPRX18AII, OnoClassA}, which leads to a catalogue of topological insulators (TIs) and topological semimetals based on a first-principle calculation for the material database.~\cite{ZhangCatalogue, VergnioryCatalogue, TangCatalogue}

In this note, we present two directions to extend the SI for other classes of topological materials. 
The one is the SIs for superconducting states. 
We see that the careful definition of topological invariants of the Bogoliubov-de Gennes (BdG) Hamiltonian relative to the trivial vacuum BdG Hamiltonian gives us new types of SIs. 
Another direction is the SIs for detecting ingap bound states (,and Andreev bound states for superconductors) localized at the boundary of a sample. 
The emergence of such ingap boundary states depends on the choice of the unit cell compatible with the boundary termination. 
Nevertheless, once we fix a unit cell, there emerges a hierarchy of the set of atomic insulators within the unit cell, giving alternative SIs associated with a small set of atomic insulators localized at the inertia of the unit cell. 

This note is inspired by some prior works. 
Ono and Watanabe discussed the implication of the SIs of electric materials for superconducting BdG Hamiltonians.~\cite{OnoSIforSC}
We show that for BdG Hamiltonians there emerge SIs to detect TSCs and gapless phases specific to superconductors (SCs). 
In preparing this work, we became aware of Ref.~\cite{SkurativskaSISC}, which also discuss the SI for SCs and gives the $\Z_{2^d}$-valued SI for $d$ space dimensions for odd parity SCs. 
Benalcazar, Li and Hughes introduced the momentum-space topological invariants for the fractional corner charge in $2d$ spinless insulators with $C_n$ rotation symmetry.~\cite{BenalcazarCorner}
Also, Ref.~\cite{SchindlerCorner} extended their strategy to spinful electrons. 

In this paper, we just argue some routes to generalize the SI. 
We leave the classification and derivation of the SIs as future works. 

The plan of this paper is as follows. 
In Sec.~\ref{sec:formulation}, we give a mathematical formulation of the SI so that it is easy to see generalizations. 
The hierarchical structure associated with various definitions of nontrivial topology of the band structure is presented. 
In Sec.~\ref{sec:exs}, we give several examples of the SIs beyond electric materials. 
We leave Appendix~\ref{app:homo} for how to compute homomorphisms $f: A \to B$ of abelian groups, which can be used to make the algorism to compute the classification of SIs for superconductors. 

\noindent
{\it Notations---}
We use the following notations. 
$\mu^{(p)}_i$: SIs for gapless states in $p$-cells in the momentum space. 
$\nu^{(p)}_i$: SIs for $p$th-order TIs/TSCs. 
$\xi^{(p)}_i$: SIs for $(p-1)$-dimensional ingap boundary states.

\section{Formulation}
\label{sec:formulation}
In this section, we illustrate how the SI is formulated for the band theory. 
To make our discussion concrete, we consider 2-dimensional systems. 
The formulation for generic space dimensions is parallel. 
In Sec.~\ref{sec:si}, we assume for simplicity the abelian group $E_1^{0,0}$ of topological invariants has no torsion. 
In cases where $E_1^{0,0}$ includes torsion, one can formulate the SI by using the method described in Appendix~\ref{app:homo}. 

\subsection{Outline}
Let $E_1^{0,0}$ be the set of abelian groups for the spaces of topological invariants at high-symmetric points in the Brillouin zone (BZ).~\footnote{
In what follows, we borrow the notation from the Aityah-Hirzebruch spectral sequence associated with the filtration of the BZ so that the $p$-skeletons are given by a cell decomposition of the BZ so that all high-symmetric subspaces are contained in a some cell. 
}
Elements of $E_1^{0,0}$ are characterized by the set $\{n_j\}$ of topological invariants at high-symmetric points. 
$n_j$ are typically given by the numbers of irreps at high-symmetric points, and take values in $\Z$ or $\Z_2$. 
Later, we see the explicit forms of SIs are written as linear combinations of $n_j$s. 

The concept of the SI would be summarized as ``something nontrivial'' that is detectable from irreps at high-symmetric points. 
We would like to exclude the group $X$ composed of ``something trivial'' from the group $E_1^{0,0}$. 
Then, the SI associated with $X$ lives in the quotient group $E_1^{0,0}/X$. 
There are many choices of the space $X$. 
The best way to consider the structure behind the SI is the sequence of subgroups 
\begin{align}
&0 \subset f^{\rm AI}_{d \leq 0}(\{ {\rm AI}_{d \leq 0}\})
\subset f^{\rm AI}_{d \leq 1}(\{ {\rm AI}_{d \leq 1}\} )
\subset f^{\rm AI}_{d \leq 2}(\{ {\rm AI}_{d \leq 2}\} )
\nonumber
\\
&\qquad 
\cong f^{\rm TI}_{d \leq 0}(\{ {\rm TI}_{d \leq 0}\})
\subset f^{\rm TI}_{d \leq 1}(\{ {\rm TI}_{d \leq 1}\})
\subset f^{\rm TI}_{d \leq 2}(\{ {\rm TI}_{d \leq 2}\} )
\nonumber 
\\
& \qquad \qquad 
\cong E_3^{0,0}
\subset E_2^{0,0}
\subset E_1^{0,0}
\label{eq:subgroup_str}
\end{align}
This is the filtration for 2-spatial dimensions. 
The ingredients above are introduced in order. 

The groups $E_p^{0,0} (p=2,3)$ are defined as the kernel of the $(p-1)$th differential of the Atiyah-Hirzebruch spectral sequence~\cite{KS_Atiyah}
\begin{align}
E_{p+1}^{0,0}:= \ker [d_p^{0,0}: E_p^{0,0} \to E_p^{p,-p+1}]. 
\end{align}
Here, the group $E_p^{p,-p+1}$ is the set of abelian groups generated by gapless Dirac points in the form $\sum_{j=1}^p k_j \g_j$ within $p$-cells. 
The differential $d_p^{0,0}$ measures the obstacle to glue the band structrue specified by a data of $E_p^{0,0}$ toghther at the whole $p$-cells. 
The first differential $d_1^{0,0}$ is easily computed by using the irreducible character, which is called the compatibility relation in the band theory.~\cite{PhysRevX.7.041069}
By design, a band structure $\bm{n} = \{n_j\} \in E_1^{0,0}$ with $\bm{n} \in E_{p}^{0,0}$ and $\bm{n} \notin E_{p+1}^{0,0}$ is a (semi)metal whose band gap closes at a $p$-cell. 
Another viewpoint is the isomorphism $E_p^{0,0}/E_{p+1}^{0,0} \cong \im d_p^{0,0}$, where $d_p^{0,0}$ expresses the creation of gapless points in $p$-cells accompanied with the band inversion of the set of irreps in $E_p^{0,0}$. 

We define the group $\{{\rm TI}_{d \leq p}\} (p=0,1,2)$ as the abelian group generated by possible TIs/TSCs supported on $d$-dimensional subregions in the real space where $d$ is less than or equal to $p$. 
In particular, the group $\{ {\rm TI}_{d \leq 0}\}$ is generated by atomic insulators. 
Elements of $\{{\rm TI}_{d \leq p}\}$ can be dependent each other, but they should be exhaustive so that it covers all TIs/TSCs with dimension less than or equal to $p$. 
The homomorphism 
\begin{align}
f^{\rm TI}_{d \leq p}: \{{\rm TI}_{d \leq p}\} \to E_1^{0,0}, \qquad 
x \mapsto f^{\rm TI}_{d\leq p}(x), 
\end{align}
is defined as the set of topological invariants at high-symmetric points for the model $x \in \{{\rm TI}_{d \leq p}\}$. 
Since $x \in \{ {\rm TI}_{d \leq p}\}$ represents an insulator/superconductor, there are no gap-closing points in the BZ, implies that $f(\{{\rm TI}_{d \leq p}\}) \subset E_3^{0,0}$ for $p\leq 2$. 
Moreover, $f(\{{\rm TI}_{d \leq 2}\}) \cong E_3^{0,0}$ holds true, since the set of irreps $\bm{n} \in E_3^{0,0}$ can be glued together in the whole BZ, implying that there exists a TI/TSC in the real space. 

The definition of the group $\{ {\rm AI}_{d \leq p}\} (p=0,1,2)$ is more involved. 
It relates to the ingap localized states localized at corners, hinges of materials.~\cite{BenalcazarScience}
We define $\{ {\rm AI}_{d \leq 2}\} (\cong \{ {\rm TI}_{d \leq 0}\})$ as the abelian group generated by all atomic insulators.~\cite{Po230,TQC}
We may further define subgroups of atomic insulators as follows. 
We define $\{ {\rm AI}_{d \leq 1}\}$ as the abelian group generated by atomic insulators whose Wyckoff positions are not located at the corners of the unit cell. 
Similarly, the group $\{ {\rm AI}_{d \leq 0}\}$ is the defined as the abelian group generated by atomic insulators whose Wyckoff positions are located at the interior of the unit cell. 
The homomorphism 
\begin{align}
f^{\rm AI}_{d \leq p}: \{{\rm AI}_{d \leq p}\} \to E_1^{0,0}, \qquad x \mapsto f^{\rm AI}_{d\leq p}(x)
\end{align}
is defined again as the set of topological invariants $n_j$ of $E_1^{0,0}$ for the model $x \in \{{\rm AI}_{d \leq p}\}$. 
Then, for example, a band structure with irreps $\bm{n}$ so that $\bm{n} \in f^{\rm AI}_{d\leq 2}(\{{\rm AI}_{d \leq 2}\})$ and $\bm{n} \notin f^{\rm AI}_{d\leq 1}(\{{\rm AI}_{d \leq 1}\})$ hosts an ingap corner state. 
We should note that the definition of the unit cell is not unique for a given magnetic space group. 
It should be fixed so that it is compatible with the real-space boundary.

\subsection{The derivation of the symmetry-based indicators}
\label{sec:si}
In this section, we see how the indicator formulas are made provided that the subgroups in (\ref{eq:subgroup_str}) of $E_1^{0,0}$ are given. 
For the purpose to illustrate the formulation, we for simplicity assume $E_1^{0,0}$ is a free abelian group so that its subgroups are also free. 
This assumption simplifies the calculation of the coimage of $d_p^{0,0}$ and the cokernel of $f^{\rm AI}_{d \leq p}, f^{\rm TI}_{d\leq p}$. 
The indicator formulas will be obtained recursively. 

\subsubsection{$E_2^{0,0} \subset E_1^{0,0}$}
As noted before, we assume $E_1^{0,0}$ is free abelian. 
Let us write 
\begin{align}
&E_1^{0,0} = \bigoplus_{j=1}^m \Z[\bm{b}_j], \\
&E_1^{1,0}=\bigoplus_{j=1}^k \Z[\bm{e}_j] \oplus \bigoplus_{j=1}^l \Z_{p_j}[\bm{f}_j]. 
\end{align}
Here, $\bm{f}_j$s are generators of torsion. 
A set of irreps of $E_1^{0,0}$ is written as $\bm{n}=\sum_{j=1}^n n_j \bm{b}_j$. 
The group $E_1^{1,0}$ represents gapless points in 1-cells. 
The first differential $d_1^{0,0}:E_1^{0,0}\to E_1^{1,0}$, the compatibility relation, is expressed by a matrix $M_{d_1^{0,0}}$ as in
\begin{align}
&d_1^{0,0} (\bm{b}_1,\dots, \bm{b}_m) = (\bm{e}_1,\dots, \bm{e}_k;\bm{f}_1,\dots,\bm{f}_l) M_{d_1^{0,0}}, \qquad 
M_{d_1^{0,0}}=
\left[
\begin{array}{c}
A \\
c_1 \\
\vdots \\
c_l \\
\end{array}
\right], \\
&A \in {\rm Mat}_{k \times m}(\Z), \qquad 
c_j \in {\rm Mat}_{1\times m}(\Z/p_j \Z), \qquad (j=1,\dots l).  
\end{align}
The group $E_2^{0,0}=\ker d_1^{0,0} \subset E_1^{0,0}$ and the quotient $E_1^{0,0}/E_2^{0,0}$ in which the SIs live are given as follows (see Appendix~\ref{app:homo}). 
First we introduce an integer-valued matrix 
\begin{align}
\tilde M_{d_1^{0,0}}=
\left[
\begin{array}{c|ccc}
A&&O\\
\hline
\tilde c_1&p_1\\
\vdots&&\ddots \\
\tilde c_l&&&p_l \\
\end{array}
\right] \in {\rm Mat}_{(k+l)\times(m+l)}(\Z), 
\end{align}
with $c_j \mapsto \tilde c_j \in {\rm Mat}_{1\times m}(\Z), (j=1,\dots l),$ integral lifts of $c_j$s.
We compute the Smith normal form (SNF) of $\tilde M_{d_1^{0,0}}$ to get 
\begin{align}
u\tilde M_{d_1^{0,0}}v = 
\left[
\begin{array}{ccc|c}
\lambda_1&&&\\
&\ddots&&O\\
&&\lambda_q&\\
\hline 
&O&&O\\
\end{array}
\right], 
\end{align}
where $\lambda_i (i=1,\dots q)$ are nonnegative integers, and $u,v$ are unimodular matrices. 
The following linear combinations of $\bm{b}_j$s generates $\ker d_1^{0,0}$, 
\begin{align}
E_2^{0,0} = \ker d_1^{0,0} = \left\langle \left\{ \sum_{i=1}^m \bm{b}_i v_{ij} \right\}_{j=q+1}^{m+l} \right\rangle.
\end{align}
To find the explicit basis of $E_2^{0,0}$, we introduce the submatrix
\begin{align}
v_{\rm sub}
:=
\left[ 
\begin{array}{cccccc}
v_{1,q+1}&\cdots&v_{1,m+l}\\
\vdots&&\vdots\\
v_{m,q+1}&\cdots&v_{m,m+l}\\
\end{array}
\right] \in {\rm Mat}_{m \times (m+l-q)}(\Z), 
\end{align}
and consider the SNF of it, 
\begin{align}
u^{(1)} v_{\rm sub} v^{(1)}
= 
\left[
\begin{array}{cc}
D^{(1)}&O\\
O&O\\
\end{array}
\right], \qquad 
D^{(1)}=
\left[
\begin{array}{ccc}
d^{(1)}_1&&\\
&\ddots&\\
&&d^{(1)}_{m_1}\\
\end{array}
\right]. 
\label{eq:snf_1}
\end{align}
With this, define new basis of $E_1^{0,0}$ by 
\begin{align}
(\bm{b}^{(1)}_1,\dots,\bm{b}^{(1)}_m)
=
(\bm{b}_1,\dots,\bm{b}_m) [u^{(1)}]^{-1}. 
\end{align}
We have the desired results 
\begin{align}
&E_2^{0,0}=\bigoplus_{j=1}^{m_1} \Z[d^{(1)}_j \bm{b}^{(1)}_i], \\
&E_1^{0,0}/E_2^{0,0}=
\bigoplus_{j=1}^{m_1} \Z_{d_j^{(1)}}[\bm{b}^{(1)}_i] \oplus 
\bigoplus_{j=m_1+1}^m \Z[\bm{b}^{(1)}_i]. 
\end{align}
The explicit formulas of the SIs are read from the set $\bm{n}$ of irreps of $E_1^{0,0}$ in the basis of $\{\bm{b}^{(1)}_j\}_{j=1}^m$. 
The SIs detecting the quotient $E_1^{0,0}/E_2^{0,0}$ are given as 
\begin{align}
\mu^{(1)}_i:= \sum_{i=1}^m [u^{(1)}]_{ij} n_j \in 
\left\{\begin{array}{ll}
\Z/d_j^{(1)} \Z & (j=1,\dots, m_1), \\
\Z & (j=m_1+1,\dots, m).
\end{array}\right.
\end{align}
Here, $d^{(1)}_j=1$ just means no indicators exist for such $j$. 
Note that $\mu^{(1)}_i$ themselves are defined as integer values, but the indicators take values in $\Z/d^{(1)}_j \Z$ or $\Z$. 
A nontrivial value of a SI $\mu_i^{(1)}\neq 0$ implies that there exists a gapless point in a $1$-cell.

\subsubsection{$E_3^{0,0} \subset E_2^{0,0}$}
In the same way as before, we have the kernel and the cockernel of the second deferential $d_2^{0,0}: E_2^{0,0} \to E_2^{2,-1}$. 
We compute the SNF of the submatrix like as in (\ref{eq:snf_1}) to get unimoduar matrices $u^{(2)}, v^{(2)}$ and the diagonal matrix $D^{(2)}$. 
Introducing the new basis of $E_2^{0,0}$ by 
\begin{align}
&(\bm{b}^{(2)}_1,\dots,\bm{b}^{(2)}_{m_1})
:=
(d_1^{(1)}\bm{b}^{(1)}_1,\dots,d^{(1)}_{m_1}\bm{b}^{(1)}_{m_1}) [u^{(2)}]^{-1}
=(\bm{b}^{(1)}_1,\dots,\bm{b}^{(1)}_{m_1}) D^{(1)} [u^{(2)}]^{-1}
,
\end{align}
we have 
\begin{align}
&E_3^{0,0}=\bigoplus_{j=1}^{m_2} \Z[d^{(2)}_j \bm{b}^{(2)}_j], \\
&E_2^{0,0}/E_3^{0,0}=
\bigoplus_{j=1}^{m_2} \Z_{d^{(2)}_j}[\bm{b}^{(2)}_j] \oplus 
\bigoplus_{j=m_2+1}^{m_1} \Z[\bm{b}^{(2)}_j]. 
\end{align}
The second SIs are given as 
\begin{align}
\mu^{(2)}_i = 
\sum_{j=1}^{m_1}[u^{(2)}]_{ij} \times \frac{\mu^{(1)}_j}{d^{(1)}_j}, \qquad  (i=1,\dots,m_1).
\end{align}
Note that the SIs $\mu^{(2)}_i$ can be fractional numbers if the first SIs $\mu_i^{(1)}$ are nonzero. 
When $\mu^{(1)}_i \equiv 0$, namely $\mu^{(1)}_i/d^{(1)}_i \in \Z$, the second SIs take values in integers as 
\begin{align}
\mu^{(2)}_i \in 
\left\{\begin{array}{ll}
\Z/d^{(2)}_j\Z & (j=1,\dots, m_2), \\
\Z & (j=m_2+1,\dots, m_1).
\end{array}\right.
\end{align}
A nontrivial value of the second SI $\mu^{(2)}_i \neq 0$ implies that there exists a gapless point in the form of $2d$ Dirac Hamiltonian $k_1\g_1+k_2\g_2$ inside a 2-cell.

\subsubsection{$\{{\rm TI}_{d\leq 1}\} \to E_1^{0,0}$}
The abelian group $E_3^{0,0}$ expresses the set of topological numbers at high-symmetric points that can extend to the entire BZ without a gapless point, i.e., $E_3^{0,0}$ represents band insulators/superconductors. 
The next step is to derive the SI for 1st-order TIs/TSCs. 
As noted before, in 2-spatial dimensions, all possible insulators/superconductors $\{{\rm TI}_{d \leq 2}\}$ cover the group $E_3^{0,0}$. 
Put differently, $f^{\rm TI}_{d \leq 2}(\{{\rm TI}_{d \leq 2}\})=E_3^{0,0}$. 

Therefore, the first SI detecting TIs/TSCs, which we denote by $\nu^{(1)}_i$, arises from the subgroup $f^{\rm TI}_{d\leq 1}(\{{\rm TI}_{d \leq 1}\}) \subset E_3^{0,0}$, where $\{{\rm TI}_{d \leq 1}\}$ is the abelian group generated by TIs/TSCs supported on 1 dimensional subspaces in the real-space manifold.~\footnote{
There is no efficient algorism to get $\{{\rm TI}_{d \leq p}\}$ yet. 
Here, we simply assume that $\{{\rm TI}_{d \leq p}\}$ is given. }
Then, by evaluating the topological invariants at high-symmetric points for models of $\{{\rm TI}_{d \leq 1}\}$, we have the homomorphism 
\begin{align}
f^{\rm TI}_{d\leq 1}: \{{\rm TI}_{d \leq 1}\} \to E_1^{0,0} 
\end{align}
in the form 
\begin{align}
f(\bm{c}_1,\dots,\bm{c}_r) = (\bm{b}_1,\dots,\bm{b}_m) M_{f^{\rm TI}_{d \leq 1}}, 
\end{align}
where $\bm{c}_j (j=1,\dots,r)$ express TIs/TSCs with the dimension less than or equal to 1. 
Note that the generators $\{\bm{c}_j\}_{j=1}^r$ include atomic insulators.  
Since we have assumed $E_1^{0,0}$ is free, $\{{\rm TI}_{d \leq 1}\}$ can be assumed to be a free abelian group. 
To derive the SI for the 1st-order TIs/TSCs, we rewrite the homomorphism $f^{\rm TI}_{d \leq 1}:\{{\rm TI}_{d \leq 1}\} \to E_1^{0,0}$ in the basis of $E_3^{0,0}$ as in 
\begin{align}
f(\bm{c}_1,\dots,\bm{c}_r) 
= 
(d^{(2)}_1\bm{b}^{(2)}_1,\dots,d^{(2)}_{m_2}\bm{b}^{(2)}_{m_2},\bm{b}^{(2)}_{m_2+1},\dots,\bm{b}^{(2)}_{m_1},\bm{b}^{(1)}_{m_1+1},\dots,\bm{b}^{(1)}_m) \tilde M_{f^{\rm TI}_{d \leq 1}}, 
\end{align}
\begin{align}
\tilde M_{f^{\rm TI}_{d \leq 1}}
=
\left[
\begin{array}{cc}
[D^{(2)}]^{-1} & \\
& I_{m-m_2} \\
\end{array}
\right]
\times 
\left[
\begin{array}{cc}
u^{(2)} [D^{(1)}]^{-1} & \\
& I_{m-m_1} \\
\end{array}
\right]
\times 
u^{(1)}
\times 
M_{f^{\rm TI}_{d\leq 1}}.
\end{align}
Since $f^{\rm TI}_{d\leq 1}(\{{\rm TI}_{d \leq 1}\}) \subset E_3^{0,0}$, the representation matrix can be written as 
\begin{align}
\tilde M_{f^{\rm TI}_{d \leq 1}}
=
\left[
\begin{array}{c}
\tilde M_{f^{\rm TI}_{d \leq 1}}^{\rm sub}\\
O\\
\end{array}
\right], \qquad 
\tilde M_{f^{\rm TI}_{d \leq 1}}^{\rm sub} \in {\rm Mat}_{m_2 \times r}(\Z). 
\end{align}
Introducing the SNF of $\tilde M_{f^{\rm TI}_{d \leq 1}}^{\rm sub}$, 
\begin{align}
u^{(3)} \tilde M_{f^{\rm TI}_{d \leq 1}}^{\rm sub} v^{(3)}
=
\left[
\begin{array}{cc}
D^{(3)}&O\\
O&O\\
\end{array}
\right], \qquad 
D^{(3)}
=
\left[
\begin{array}{ccc}
d_1^{(3)}&&\\
&\ddots&\\
&&d^{(3)}_{m_3}\\
\end{array}
\right], 
\label{eq:snf_3}
\end{align}
we have 
\begin{align}
f(\bm{c}_1,\dots,\bm{c}_r)v^{(3)} 
= 
(\bm{b}^{(3)}_1,\dots,\bm{b}^{(3)}_{m_2},\bm{b}^{(2)}_{m_2+1},\dots,\bm{b}^{(2)}_{m_1},\bm{b}^{(1)}_{m_1+1},\dots,\bm{b}^{(1)}_m)
\left[
\begin{array}{ccc|c}
d_1^{(3)}&&&\\
&\ddots&&O\\
&&d^{(3)}_{m_3}&\\
\hline
&O&&O\\
\end{array}
\right], 
\end{align}
with 
\begin{align}
(\bm{b}^{(3)}_1,\dots,\bm{b}^{(3)}_{m_2})
:=
(\bm{b}^{(2)}_1,\dots,\bm{b}^{(2)}_{m_2}) [u^{(3)}]^{-1}. 
\end{align}
The SIs to detect the 1st-order TIs/TSCs are given as  
\begin{align}
\nu^{(1)}_i
=
\sum_{j=1}^{m_2}[u^{(3)}]_{ij} \times \frac{\mu^{(2)}_j}{d^{(2)}_j}
\in 
\left\{\begin{array}{ll}
\Z/d^{(3)}_j\Z & (j=1,\dots, m_3), \\
\Z & (j=m_3+1,\dots, m_2).
\end{array}\right.
\end{align}
Again, the SI $\nu^{(1)}_i$ can be fractional if $\mu^{(2)}_i$s have nontrivial.  
When, $\mu^{(2)}_i$ take trivial values, a nontrivial SI $\nu^{(1)}_i \neq 0$ implies a 1st-order TI/TSC.

\subsubsection{$\{{\rm AI}_{d\leq p}\} \to E_1^{0,0}$}
Let $\bm{a}_1,\dots, \bm{a}_{p_0}$ be atomic insulators located at in the inertia of the unit cell, $\bm{a}_{p_0+1},\dots,\bm{a}_{p_1}$ be those located at the edge of the unit cell not including the corner, and $\bm{a}_{p_1+1},\dots, \bm{a}_{p_2}$ be those located at the corner of the unit cell, so that they generate the groups of atomic insulators $\{{\rm AI}_{d \leq 0}\}, \{{\rm AI}_{d \leq 1}\}$ and  $\{{\rm AI}_{d \leq 2}\}$, respectively. 
In the same way as in (\ref{eq:snf_3}), we have the SNF of the homomorphisms 
\begin{align}
f^{\rm AI}_{d \leq p}: \{{\rm AI}_{d \leq p}\} \to E_1^{0,0}, 
\end{align}
where $f^{\rm AI}_{d \leq p}$ is defined by the set of topological invariants of $E_1^{0,0}$ for a given model $x \in \{{\rm AI}_{d \leq p}\}$. 

Let $u^{(4)}, v^{(4)}, D^{(4)} = {\rm diag}(d^{(4)}_1,\dots,d^{(4)}_{m_4})$ be the data of the SNF of $f^{\rm AI}_{d \leq 2}$ as in (\ref{eq:snf_3}), we have new SIs 
\begin{align}
\nu^{(2)}_i
=
\sum_{j=1}^{m_3}[u^{(4)}]_{ij} \times \frac{\mu^{(3)}_j}{d^{(3)}_j}
\in 
\left\{\begin{array}{ll}
\Z/d^{(4)}_j\Z & (j=1,\dots, m_4), \\
\Z & (j=m_4+1,\dots, m_3).
\end{array}\right.
\end{align}
Under the condition that the lower SIs $\nu^{(1)}_i, \mu^{(2)}_i, \mu^{(1)}_i$ are trivial, a nontrivial value of $\nu^{(2)}_i$ means that the band structure is a 2nd-order TI/TSC. 

Similarly, let $u^{(5)}, v^{(5)}, D^{(5)} = {\rm diag}(d^{(5)}_1,\dots,d^{(5)}_{m_5})$ be the data of the SNF of $f^{\rm AI}_{d \leq 1}$ as in (\ref{eq:snf_3}), we have the SIs to detect ingap corner states 
\begin{align}
\xi^{(1)}_i
=
\sum_{j=1}^{m_4}[u^{(5)}]_{ij} \times \frac{\mu^{(4)}_j}{d^{(4)}_j}
\in 
\left\{\begin{array}{ll}
\Z/d^{(5)}_j\Z & (j=1,\dots, m_5), \\
\Z & (j=m_5+1,\dots, m_4).
\end{array}\right.
\end{align}

At last, let $u^{(6)}, v^{(6)}, D^{(6)} = {\rm diag}(d^{(6)}_1,\dots,d^{(6)}_{m_6})$ be the data of the SNF of $f^{\rm AI}_{d \leq 0}$ as in (\ref{eq:snf_3}), we have the SIs to detect ingap edge states 
\begin{align}
\xi^{(2)}_i
=
\sum_{j=1}^{m_5}[u^{(6)}]_{ij} \times \frac{\mu^{(5)}_j}{d^{(5)}_j}
\in 
\left\{\begin{array}{ll}
\Z/d^{(6)}_j\Z & (j=1,\dots, m_6), \\
\Z & (j=m_6+1,\dots, m_5).
\end{array}\right.
\end{align}

SIs for 2-spatial dimensions based on the subgroups (\ref{eq:subgroup_str}) are summarized in Table~\ref{tab:si}.

\begin{table}
\caption{The list of SIs for two spatial dimensions.}
\label{tab:si}
\centering
\begin{align*}
\begin{array}{llllll}
\mbox{SI}&\mbox{Range}&\mbox{Target}\\
\hline
\xi^{(2)}_j (j=1,\dots,m_6)& \Z/d^{(6)}_j\Z & \mbox{Ingap edge states}\\
\xi^{(2)}_j (j=m_6+1,\dots,m_5)& \Z & \\
\hline
\xi^{(1)}_j (j=1,\dots,m_5)& \Z/d^{(5)}_j\Z & \mbox{Ingap corner states}\\
\xi^{(1)}_j (j=m_5+1,\dots,m_4)& \Z & \\
\hline
\nu^{(2)}_j (j=1,\dots,m_4)& \Z/d^{(4)}_j\Z & \mbox{2st-order TIs/TSCs}\\
\nu^{(2)}_j (j=m_4+1,\dots,m_3)& \Z & \\
\hline
\nu^{(1)}_j (j=1,\dots,m_3)& \Z/d^{(3)}_j\Z & \mbox{1st-order TIs/TSCs}\\
\nu^{(1)}_j (j=m_3+1,\dots,m_2)& \Z & \\
\hline
\mu^{(2)}_j (j=1,\dots,m_2)& \Z/d^{(2)}_j\Z & \mbox{Gapless states in 2-cells}\\
\mu^{(2)}_j (j=m_2+1,\dots,m_1)& \Z & \\
\hline
\mu^{(1)}_j (j=1,\dots,m_1)& \Z/d^{(1)}_j\Z & \mbox{Gapless states in 1-cells}\\
\mu^{(1)}_j (j=m_1+1,\dots,m)& \Z & \\
\end{array}
\end{align*}
\end{table}

\subsection{Superconductors}
\label{sec:sc}
\subsubsection{Vacuum and triple}
For SCs, we should be careful about what a nontrivial Hamiltonian is. 
Let $E$ be the one-particle Nambu-Hirbert space compsoed of an atomic insulator and its particle-hole pair on which the BdG Hamilotnian defined. 
The $K$-group is represented by a Karoubi's triple $[E,H,H_0]$ of Hamiltonians $H,H_0$ that act on the common Nambu-Hilbert space $E$.~\cite{karoubi2008k}
For SCs, we have a canonical reference Hamiltonian $H_0$: We can set $H_0$ to be the vacuum Hamiltonian for the atomic insulators 
\begin{align}
H_0(\bk) = \frac{\varepsilon}{2} \tau_z 
\end{align}
with the positive chemical potential $\epsilon>0$, which can also be written as 
\begin{align}
\hat H_0 = \varepsilon \sum_{\bm{R},\alpha,j}\hat f^\dag_{\bm{R}\alpha j} \hat f_{\bm{R}\alpha j} 
\end{align}
in the many-body Hilbert space, where $\bm{R},\alpha,j$ run over all the degrees of freedom, namely the positions of the unit cell, Wyckoff poisitoins, and internal degrees of freedom, respectively, and $\hat f^\dag_{\bm{R}\alpha j}, \hat f_{\bm{R}\alpha j}$ are complex fermion creation and annihilation operators. 
The positive chemical potential means that the ground state of $\hat H_0$ is the vacuum state $\ket{0}$ of the complex fermions. 
In other words, for the SI, we should define topological invariants at high-symmetric points for a given BdG Hamiltonian $H$ as a relative index of the pair $[H,H_0]$ with $H_0=\tau_z$ the reference BdG Hamiltonian.

\subsubsection{Atomic insulators}
We also define atomic insulators for superconductors. 
They are defined as the atomic insulators in the usual sense. 
For a given Nambu-Hilbert space $E$, the atomic insulator is defined as the fully occupied state that is represented by the BdG Hamiltonian $H(\bk) = \frac{\epsilon}{2} \tau_z$ with a negative chemical potential $\epsilon <0$. 
Therefore, as an element of the $K$-group, the atomic insulator for a Nambu-Hilbert space $E$ is given by the triple $[E,H=-\tau_z,H_0=\tau_z]$. 

\subsubsection{Weak coupling limit}
In usual, the superconducting gap function $\Delta(\bk)$ is much smaller than the energy scale of the normal state $h(\bk)$. 
This means that topological invariants of the BdG Hamiltonian  
\begin{align}
H=\begin{pmatrix}
h(\bk)&\Delta(\bk)\\
\Delta(\bk)^\dag&-h(-\bk)^T\\
\end{pmatrix}
\end{align}
at high-symmetric points, which compose SIs, can be usually computed solely by the normal part $h(\bk)$ by adiabatically decreasing the gap function $\Delta(\bk)$ to zero. 
We call this simplification of the topological invariants the weak coupling limit.

\section{Some examples}
\label{sec:exs}
In this section, we illustrate the framework developed in Sec.~\ref{sec:formulation} with several examples. 
Some parts of this section are intentionally written long. 
It is for the purpose to be easily extensible to general magnetic space groups.

\subsection{$1d$ SCs}
\label{sec:1dsc}
Let us start with the simplest example of the SI for superconductors, that is, $1d$ class D systems with only translation symmetry. 
The symmetry constraint in the $k$-space is just the class D PHS 
\begin{align}
CH(k)C^{-1}=-H(-k), \qquad C=\tau_x K, 
\label{eq:1dsc_phs}
\end{align}
where $K$ is the complex conjugation.
At two high-symmetric points $k=0,\pi$ the effective AZ class are class D, while that for generic points is class A. 
Thus, we have $E_1^{0,0}=\Z_2+\Z_2$ and $E_1^{1,0}=\Z$. 
Here, $\Z_2$ is generated by the triple 
\begin{align}
[E=\C^2, C=\tau_xK, H=-\tau_z,H_0=\tau_z], 
\label{eq:1dsc_z2_generator}
\end{align}
and characterized by the $\Z_2$-quantized Pfaffian 
\begin{align}
n^{k} = \frac{1}{\pi} {\rm Arg} \frac{\pf [H\tau_x]}{\pf [H_0 \tau_x]} \in \{0,1\} 
\end{align} 
for $k=0,\pi$. 
In the weak coupling limit, this is rewritten as 
\begin{align}
n^{k} 
= \frac{1}{\pi} {\rm Arg} \frac{\det [h]}{\det[{\bf 1}]} 
= N[h] \quad \mod 2, 
\end{align}
where $N[H]$ denotes the number of occupied state of the Hamiltonian $H$, and $h$ is the normal part of the BdG Hamiltonian $H$. 
Similarly, $\Z$ is generated by the triple
\begin{align}
[E=\C, H=-1,H_0=1]
\end{align}
at a generic point $k \in (0,\pi)$, and characterized by the integer-valued invariant 
\begin{align}
n^{0\to \pi}=N[H_1]-N[H_0] \in \Z.
\label{eq:1dsc_zinv}
\end{align}
We also have the particle-hole symmetric pair $[E=\C,H=1,H_0=-1]$ at $-k \in (-\pi,0)$.
The first differential $d_1^{0,0}: E_1^{0,0} \to E_1^{1,0}$ is trivial~\footnote{
Since no nontrivial homomorphisms exist from a torsion $\Z_n$ to the free abelian group $\Z$. 
In fact, the $\Z$ topological invariant (\ref{eq:1dsc_zinv}) is zero for the pair $(H,H_0)$ of the generator (\ref{eq:1dsc_z2_generator}). 
}, we have $E_2^{0,0}=E_1^{0,0}$. 

The group $\{{\rm AI}_{d \leq 1}\}$ of atomic insulators is generated by the relative difference between the atomic insulator and the vacuum in the one-band system 
\begin{align}
[E=S^1 \times \C^2,C=\tau_x K, H(k)=-\tau_z,H_0(k)=\tau_z].
\end{align}
This generates the group $\{{\rm AI}_{d \leq 1}\}=\Z_2$. 
Therefore, there is a nontrivial SI associated with the quotient $E_2^{0,0}/\{{\rm AI}_{d \leq 1}\} = \Z_2$ that is characterized by the sum of $\Z_2$ invariants of $E_1^{0,0}$ at $k=0,\pi$, 
\begin{align}
\nu=n^0-n^\pi=\frac{1}{\pi} {\rm Arg} \frac{\pf [H(k=0)\tau_x]}{\pf [H(k=\pi)\tau_x]} \in \{0,1\}. 
\end{align}
This is nothing but the Pfaffian formula for the $1d$ TSC by Kitaev.~\cite{kitaev2001unpaired}
In the weak coupling limit, the above SI is simplified as 
\begin{align}
\nu
=
N[h(k=0)]-N[h(k=\pi)] \quad \mod 2. 
\label{eq:1dsc_si_wcl}
\end{align}
This means that if the normal state has an odd number of fermi points in between the momentum $k=0$ and $k=\pi$ and the superconducting order induces a mass gap to the fermi points, the system becomes a TSC.

\subsection{$1d$ odd-parity SCs}
\label{sec:1d_sc_inv_odd}
Let us consider $1d$ TRS-broken odd-parity SCs. 
The symmetry constraint is summarized as 
\begin{align}
&P(k)H(k)P(k)^{-1}=H(-k), \qquad P(-k)P(k)=1, \label{eq:1dsc_sym_inv} \\
&CP(k)=-P(-k)C.
\end{align}
in addition to the PHS (\ref{eq:1dsc_phs}). 
Here, 
\begin{align}
P(k)
=
\begin{pmatrix}
p(k)&\\
&-p(-k)^*
\end{pmatrix}_\tau
\end{align}
is the inversion operator for the BdG Hamiltonian with $p(k)$ one for the normal state. 

At the high-symmetric points $k=0,\pi$, the PHS operator $C$ exchanges two irreps $P(k)=\pm 1$, implying that the effective AZ classes are class A. 
The group $E_1^{0,0}$ is given by $E_1^{0,0} = \Z[\bm{b}^0] \oplus \Z[\bm{b}^\pi]$ where each $\Z$ is generated by the following triple 
\begin{align}
\left[
E=\C^2, C=\tau_xK, P=\tau_z, H=-\tau_z, H_0=\tau_z
\right]
\end{align}
characterized by the $\Z$ invariant 
\begin{align}
n^k=N_+[H(k)]-N_+[H_0(k)] \in \Z, 
\label{eq:1d_TRB_SC_inv}
\end{align}
with $N_{\pm}[H]$ the number of occupied states of the Hamiltonian $H$ with the positive/negative parity $P(k)=\pm$ at the high-symmetric point $k \in \{0,\pi\}$.
In the weak coupling limit, this is simplified as the difference of the numbers of positive and negative parity eigenstates in the occupied states, 
\begin{align}
n^k
=\big\{ N_+[h(k)]+N_-[-h(k)] \big\}-\big\{ N_+[{\bf 1}]+N_-[-{\bf 1}] \big\}
=N_+[h(k)]-N_-[h(k)]
\in \Z. 
\end{align}

At a generic point $k \in (0,\pi)$ in the BZ, the little group $G_k$ of symmetry group is $\Z_2$ that is generated by $CP(k)$. 
This symmetry operator satisfies $(CP(k))^2 = -1$, meaning that the effective AZ class is class C. 
No stable point nodes exist in the 1-cell. 
Therefore, $E_1^{1,0}=0$, and we have $E_2^{0,0}=E_1^{0,0}$. 

Let $\{ {\rm AI}_{d \leq 1} \}$ is the group generated by atomic insulators. 
We have two atomic insulators $\bm{a}^0, \bm{a}^\frac{1}{2}$ located at Wyckoff positions $x=0,\frac{1}{2}$ in the unit cell. 
These are classified by $\Z$ as well as that for BdG Hamiltonians in the $k$-space at a high-symmetric point, thus, $\{{\rm AI}_{d\leq 1}\}=\Z^{\oplus 2}$. 
In the $k$-space, the BdG Hamiltonians and symmetry operators for these atomic insulators and the corresponding vacuum are written as 
\begin{align}
&\bm{a}^0=
[E=S^1 \times \C^2, C=\tau_xK, P(k)=\tau_z, H(k) =-\tau_z, H_0(k)=\tau_z], \\
&\bm{a}^\frac{1}{2}=
[E=S^1 \times \C^2, C=\tau_xK, P(k)=\tau_ze^{-ik}, H(k) =-\tau_z, H_0(k)=\tau_z].
\end{align}
For these triples, the $\Z$ invariants of $E_1^{0,0}$ are computed as follows. 
\begin{align}
\begin{array}{c|cc}
&n^0&n^\pi\\
\hline
\bm{a}^0&1&1\\
\bm{a}^\frac{1}{2}&1&-1\\
\end{array}
=: (M_{f^{\rm AI}_{d \leq 1}})^T. 
\end{align}
This gives the homomorphism $f^{\rm AI}_{d \leq 1}: \{{\rm AI}_{d \leq 1}\} \to E_1^{0,0}$, 
\begin{align}
f^{\rm AI}_{d \leq 1}(\bm{a}^0,\bm{a}^\frac{1}{2})
=(\bm{b}^0,\bm{b}^\pi) M_{f^{\rm AI}_{d \leq 1}}.
\end{align}
We have the nontrivial quotient $E_1^{0,0}/\im f^{\rm AI}_{d \leq 1} =\Z/2\Z$ characterized by the SI 
\begin{align}
\nu
=n^0-n^\pi
=
\big\{N_+[H(0)]-N_+[H(\pi)]\big\}
-
\big\{N_+[H_0(0)]-N_+[H_0(\pi)]\big\}
\qquad {\rm mod\ }2.
\end{align}
The SI $\nu$ detects the same Kitaev chain phase as in Sec.~\ref{sec:1dsc}. 
In the weak coupling limit, the SI is simplified as the SI (\ref{eq:1dsc_si_wcl}) without inversion symmetry 
\begin{align}
\nu
=
\big\{ N_+[h(0)]-N_-[h(0)] \big\}-\big\{ N_+[h(\pi)]-N_-[h(\pi)] \big\}
=
N[h(0)]-N[h(\pi)] 
\qquad {\rm mod\ }2.
\end{align}

A simple example is the spinless $p$-wave SC (the Kitaev chain) 
\begin{align}
H(k)
=(-t\cos k -\mu)\tau_z + \Delta \sin k \tau_y, \qquad 
C=\tau_xK, \qquad 
P(k)=\tau_z.
\label{eq:1d_kitaev}
\end{align}
In the parameter region $|\mu|<|t|$ so that there is a fermi point in $k \in (0,\pi)$, a Majorana zero mode $\hat \g$ appears at the edge. 

In the presence of inversion symmetry, one can further construct the SI for the Andreev bound states. 
Let $\{{\rm AI}_{d\leq 0}\}=\Z[\bm{a}^0]$ be the group generated by the atomic insulator at the center of the unit cell.
The homomorphism $f^{\rm AI}_{d\leq 0}: \{{\rm AI}_{d\leq 0}\} \to E_1^{0,0}$ is given by 
\begin{align}
f^{\rm AI}_{d \leq 0}(\bm{a}^0)
=(\bm{b}^0,\bm{b}^\pi) M_{f^{\rm AI}_{d \leq 0}}, \qquad 
(M_{f^{\rm AI}_{d \leq 0}})^T=\begin{array}{c|cc}
&n^0&n^\pi\\
\hline
\bm{a}^0&1&1\\
\end{array}.
\end{align}
We have the quotient $\im f^{\rm AI}_{d\leq 1}/\im f^{\rm d\leq 0}=\Z$ characterized by the SI 
\begin{align}
\xi
:=
\frac{n^0-n^\pi}{2} \in \frac{1}{2} \Z.
\end{align} 
This takes a value in $\Z$ provided that the SI $\nu$ for the TSC is trivial. 


The SI $\xi$ for the Andreev bound state is demonstrated for the Hamiltonian $H(k)^{\oplus 2m}$ with an even number of copies of the Kitaev chains (\ref{eq:1d_kitaev}). 
The system is trivial as a TSC as the SI $\nu$ is, however, the SI $\xi$ for the Andreev bound states takes a nontrivial value $\xi=m$. 
In fact, at the edge, there exist 2m Majorana fermions and they form $m$ complex fermions with finite energy.

\subsection{$1d$ even-parity SCs}
Let us consider $1d$ TRS-broken even-parity SCs. 
We have the same symmetry constraints as (\ref{eq:1dsc_phs}) and (\ref{eq:1dsc_sym_inv}), but the different algebra for the inversion and PHS operators 
\begin{align}
&CP(k)=P(-k)C 
\end{align}
with 
\begin{align}
P(k)
=
\begin{pmatrix}
p(k)&\\
&p(-k)^*
\end{pmatrix}_\tau. 
\end{align}

The group $E_1^{0,0}$ is given by $E_1^{0,0} = \Z_2[\bm{b}^0_+]\oplus \Z_2[\bm{b}^0_-]\oplus\Z_2[\bm{b}^\pi_+]\oplus\Z_2[\bm{b}^\pi_-]$, where each $\bm{b}^k_\pm$ is generated by the triple 
\begin{align}
\bm{b}^k_\pm
=\left[
\{k\} \times \C^2,C=\tau_xK, P=\pm, H=-\tau_z, H_0=\tau_z
\right]
\end{align}
that is characterized by the $\Z_2$-quantized Pfaffian  
\begin{align}
n^k_\pm 
=
\frac{1}{\pi} {\rm Arg} \frac{\pf \left[\frac{1\pm P(k)}{2}H(k)\tau_x\right]}{\pf \left[\frac{1\pm P(k)}{2}H_0(k)\tau_x\right]} \in \{0,1\}, 
\end{align}
where $\frac{1\pm P(k)}{2}, (k =0,\pi ),$ is the projector onto the positive/negative parity states. 
In the weak coupling limit, $n^k_\pm$ is reduced as 
\begin{align}
n^k_\pm
&=N_\pm[h(k)] \quad \mod 2, \qquad k \in \{0,\pi\}.
\end{align}

At a generic point $k\in (0,\pi)$, the little group is $\Z_2$ generated by $CP(k)$ with $(CP(k))^2=1$. 
The effective AZ class is class D, and we have $E_1^{1,0} = \Z_2[\bm{b}^{0 \to \pi}]$ with the generator
\begin{align}
\bm{b}^{0\to \pi}
= \left[
\{k \in (0,\pi)\} \times \C^2,CP(k)=\tau_xK, H=-\tau_z, H_0=\tau_z
\right] 
\end{align}
characterized by the Pfaffian 
\begin{align}
n^{0\to \pi}
=
\frac{1}{\pi} {\rm Arg} \frac{\pf \left[H(k)\tau_xP(k)^* \right]}{\pf \left[H_0(k)\tau_xP(k)^*\right]} \in \{0,1\}, \quad k\in (0,\pi), 
\end{align}
and 
\begin{align}
n^{0\to \pi}
=N[h(k)]\quad \mod 2, \quad k\in (0,\pi),
\end{align}
in the weak coupling limit. 

The compatibility relation, which is the first differential $d_1^{0,0}: E_1^{0,0} \to E_1^{1,0}$, is given by 
\begin{align}
d_1^{0,0}(\bm{b}^0_+,\bm{b}^0_-,\bm{b}^\pi_+,\bm{b}^\pi_-)
=
\bm{b}^{0 \to \pi} \ 
\left[
\begin{array}{cccc}
1&1&1&1\\
\end{array}
\right].
\end{align}
We have a nontrivial quotient $E_1^{0,0}/E_2^{0,0} = \Z_2$ charactrized by the SI 
\begin{align}
\mu = n^0_++n^0_-+n^\pi_++n^\pi_- \quad \mod 2, 
\end{align}
which is reduced to 
\begin{align}
\mu
=
N[h(0)]-N[h(\pi)] \quad \mod 2
\end{align}
in the weak coupling limit. 
When $\mu = 1 \mod 2$, we have a nodal point (Bogoliubov Fermi surface) in a 1-cell, irrespective of the gap function $\Delta(k)$. 

A simple example is the spinless SC
\begin{align}
H(k) =(-t \cos k -\mu) \tau_z + \re \Delta(k)\tau_x-\im \Delta(k)\tau_y, 
\qquad |\mu| < |t|, 
\qquad 
C=\tau_x K. 
\end{align}
When the gap function $\Delta(k)$ obeys the even-parity condition $\Delta(-k)=\Delta(k)$, one can show $\Delta(k)$ vanishes, implies that the fermi point is stable. 

The next step is to compute the quotient $E_2^{0,0}/\im f^{\rm AI}_{d\leq 1}$. 
It turns out to be zero, which can proven as follows. 
One can show that there is no $1d$ TCs compatible with the inversion symmetry with even-parity gap function, thus, we have $\{{\rm TI}_{d\leq 1}\}=0$. 
Therefore, the subgroup of $E_2^{0,0}$ starts at the group $\{{\rm AI}_{d\leq 1}\}$ of $1d$ atomic insulators, and so the homomorphism $f^{\rm AI}_{d\leq 1}: \{{\rm AI}_{d\leq 1}\} \to E_2^{0,0}$ is surjective.

In a similar way to Sec.~\ref{sec:1d_sc_inv_odd}, one can construct the SI for Andreev bound states. 
Let $\{{\rm AI}_{d\leq 0}\} = \Z_2[\bm{a}^0_+]\oplus \Z_2[\bm{a}^0_-]$ be the group generated by the atomic insulators localized at the center of the unit cell. 
Two generators are given by the triples 
\begin{align}
\bm{a}^0_\pm
=
[S^1 \times \C^2,C=\tau_xK, P(k)=\pm, H(k)=-\tau_z, H_0(k)=\tau_z], 
\end{align}
and have topological invariants 
\begin{align}
\begin{array}{c|cccc}
&n^0_+&n^0_-&n^\pi_+&n^\pi_-\\
\hline
\bm{a}^0_+&1&0&1&0\\
\bm{a}^0_-&0&1&0&1\\
\end{array}
=: (M_{f^{\rm AI}_{d\leq 0}})^T. 
\end{align}
This gives the homomorphism $f^{\rm AI}_{d\leq 0}: \{{\rm AI}_{d\leq 0}\} \to E_1^{0,0}, 
f^{\rm AI}_{d=0}(\bm{a}^0,\bm{a}^\frac{1}{2})
=(\bm{b}^0_+,\bm{b}^0_-,\bm{b}^\pi,\bm{b}^\pi_-) M_{f^{\rm AI}_{d=0}}$.
We have the nontrivial quotient $E_2^{0,0}/\im f^{\rm AI}_{d\leq 0} =\Z_2$ with the SI for the Andreev bound state
\begin{align}
\xi
=n^0_-+n^\pi_- \quad \mod 2. 
\end{align}
This is recast as 
\begin{align}
\xi
=N_-[h(0)]+N_-[h(\pi)] \quad \mod 2 
\end{align}
in the weak coupling limit. 
Under the assumption of the triviality of the SI $\mu$ and that the dof are not located at the boundary of the unit cell, $\xi \neq 0$ implies the existence of an ingap edge state of the BdG Hamiltonian, i.e., the Andreev bound state. 

A prime example is the two-orbital SC with an even-parity inter-orbital gap function (equivalently, the spinful SC with an odd-mirror gap function). 
\begin{align}
H(k)
=(-t\cos k -\mu)\tau_z + \Delta \sin k \tau_y \s_x, \qquad 
C=\tau_xK, \qquad 
P(k)=\s_z.
\end{align}
At the edge there are two Majoran zero modes $\hat \g_{\s_x=+}, \hat \g_{\s_x=-}$, and 
these Majorana fermions form a complex fermion $\psi=(\hat \g_{\s_x=+}+i \hat \g_{\s_x=-})/2$, and it may have a finite energy $\epsilon \psi^\dag \psi$ depending on the microscopic structure of the edge.

\subsection{$2d$ TR-symmetric spinless systems with $C_4$ rotation symmetry}

\begin{figure}[!]
\centering
\includegraphics[width=\linewidth, trim=0cm 2cm 0cm 3cm]{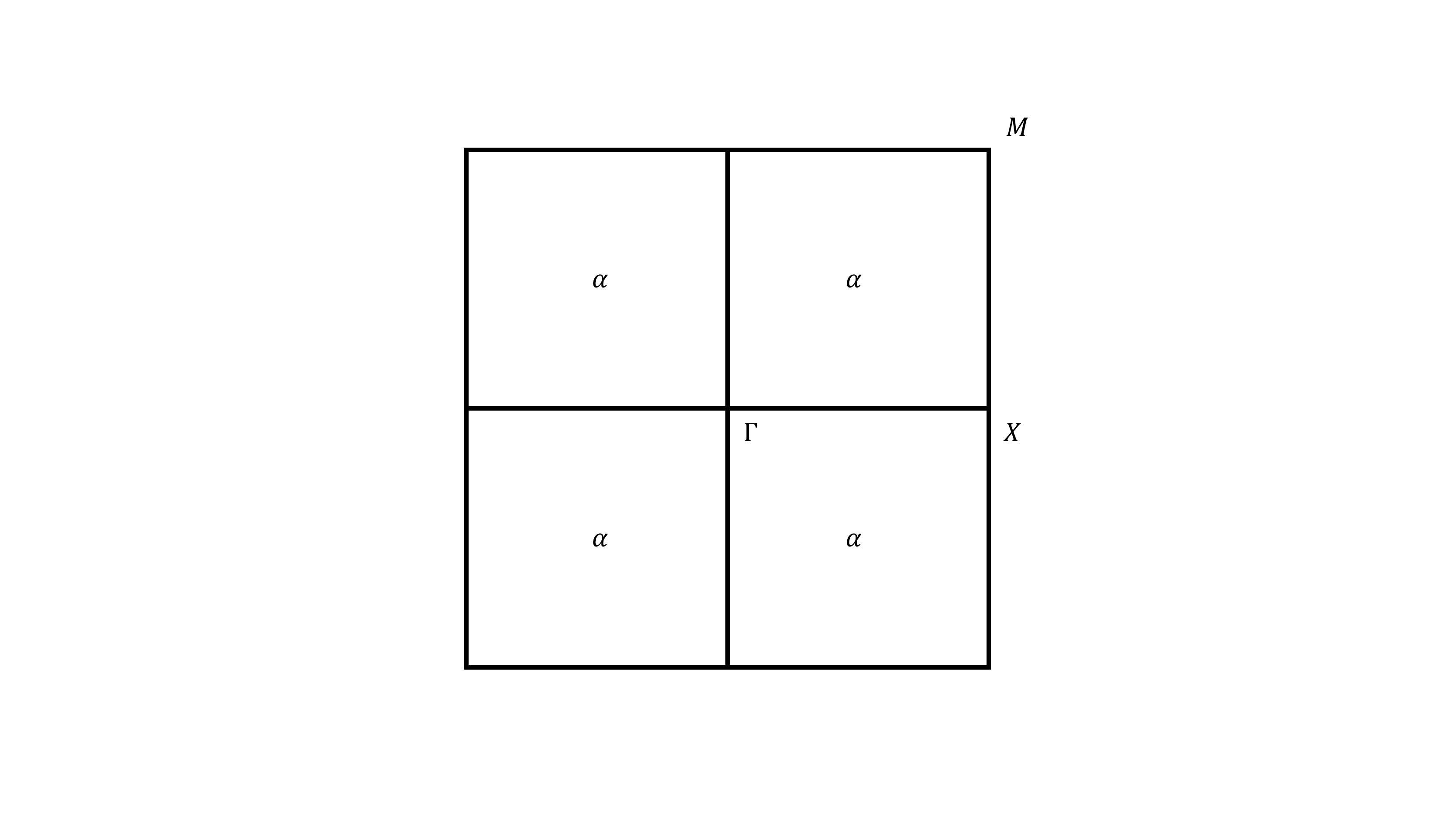}
\caption{A $C_4$ symmetric $1$-skeleton of the $2d$ BZ.}
\label{fig:c4}
\end{figure}

Let us consider $2d$ spinless electrons with time-reversal (TR) and $C_4$-rotation symmetry. 
The relationship between the corner state and the momentum-space invariants in this system was examined in Ref.~\cite{BenalcazarCorner}.
The symmetry constraint and the algebra among the symmetry operators in the momentum space is summarized as 
\begin{align}
&TH(\bk)T^{-1}=H(-\bk), \qquad T^2=1, \nonumber \\
&C_4(\bk)H(k_x,k_y)C_4(\bk)^{-1}=H(c_4\bk), \qquad  C_4(c_4^3\bk)C_4(c_4^2\bk)C_4(c_4\bk)C_4(\bk)=1, \nonumber \\
&TC_4(\bk)=C_4(-\bk)T, 
\end{align}
where $c_4\bk=(-k_y,k_x)$. 
At generic points in the BZ, the effective AZ class is AI, where effective TRS is the composition $TC_2(\bk)$ with $(TC_2(\bk))^2=1$. 
The group $E_1^{0,0}$ is given by $E_1^{0,0}=\Z^3+\Z^3+\Z^2$ generated respectively by irreps at $\G$, $M$, and $X$ points. 
We employ the 1-skeleton of the BZ as shown in Fig.~\ref{fig:c4}. 
The group $E_1^{1,0}=\Z+\Z$ is composed of irreps at 1-cells $\G \to X$ and $X \to M$.  
Let 
\begin{align}
(\bm{b}^\G_1,\bm{b}^\G_{-1},\bm{b}^\G_{(i,-i)},\bm{b}^M_1,\bm{b}^M_{-1},\bm{b}^M_{(i,-i)},\bm{b}^X_1,\bm{b}^X_{-1}) 
\end{align}
be the basis of $E_1^{0,0}$, where the superscripts and subscripts represent high-symmetric points and irreps, respectively. 
Similally, let $(\bm{c}^{\G\to X},\bm{c}^{X\to M})$ be the basis of $E_1^{1,0}$. 

\subsubsection{$E_1^{0,0} \to E_1^{1,0}$}
The compatibility relation define the first differential 
\begin{align}
d_1^{0,0}: E_1^{0,0} \to E_1^{1,0}, \qquad 
d_1^{0,0}(\bm{b}^\G_1,\dots, \bm{b}^X_{-1})=(\bm{c}^{\G\to X},\bm{c}^{X\to M})M_{d_1^{0,0}}, 
\end{align}
\begin{align}
M_{d_1^{0,0}}
=
\begin{array}{ccc|ccc|cc|l}
\bm{b}^\G_1&\bm{b}^\G_{-1}&\bm{b}^\G_{(i,-i)}&\bm{b}^M_1&\bm{b}^M_{-1}&\bm{b}^M_{(i,-i)}&\bm{b}^X_1&\bm{b}^X_{-1}\\
\hline
1&1&2&&&&-1&-1&\bm{c}^{\G\to X}_1\\
&&&-1&-1&-2&1&1&\bm{c}^{X\to M}_{-1}\\
\end{array}. 
\end{align}
The SNF of $M_{d_1^{0,0}}$ is given as 
\begin{align}
uM_{d_1^{0,0}}v
= 
\left[
\begin{array}{cc}
I_2 & O \\
\end{array}
\right], 
\end{align}
where $u,v$ are unimodular. 
Introduce the submatrix $v_{\rm sub}=\{ v_{ij} \}_{1\leq i\leq 8, 3\leq j \leq 8}$ that spans $\ker d_1^{0,0}$. 
The SNF of $v_{\rm sub}$ gives the basis of $E_2^{0,0}$ and the SIs. 
We have 
\begin{align}
u^{(1)} v_{\rm sub} v^{(1)}
=
\left[
\begin{array}{c}
I_6\\
O
\end{array}
\right], 
\end{align}
and $E_2^{0,0}=\bigoplus_{j=1}^6 \Z \bm{b}^{(1)}_j$ with $(\bm{b}^{(1)}_1,\dots, \bm{b}^{(1)}_8)=(\bm{b}^\G_1,\dots, \bm{b}^X_{-1})[u^{(1)}]^{-1}$, 
\begin{align}
[u^{(1)}]^{-1}=
\begin{array}{cccccc|cc|l}
\bm{b}^{(1)}_1&\bm{b}^{(1)}_2&\bm{b}^{(1)}_3&\bm{b}^{(1)}_4&\bm{b}^{(1)}_5&\bm{b}^{(1)}_6&\bm{b}^{(1)}_7&\bm{b}^{(1)}_8&\\
\hline
 -2 & -1 & 0 & 0 & 1 & 1 & 1 & 0&\bm{b}^\G_1\\
 0 & 1 & 0 & 0 & 0 & 0 & 0 & 0 &\bm{b}^\G_{-1}\\
 1 & 0 & 0 & 0 & 0 & 0 & 0 & 0 &\bm{b}^\G_{(i,-i)}\\
 0 & 0 & -1 & -2 & 1 & 1 & 0 & 0 &\bm{b}^M_1\\
 0 & 0 & 1 & 0 & 0 & 0 & 0 & 0 &\bm{b}^M_{-1}\\
 0 & 0 & 0 & 1 & 0 & 0 & 0 & 0 &\bm{b}^M_{(i,-i)}\\
 0 & 0 & 0 & 0 & 1 & 0 & 0 & 0 &\bm{b}^X_1\\
 0 & 0 & 0 & 0 & 0 & 1 & 0 & 1 &\bm{b}^X_{-1}\\
\end{array}. 
\end{align}
We get two $\Z$-valued SIs $\mu^{(1)}_7, \mu^{(1)}_8$ to detect the quotient group $E_2^{0,0}/E_1^{0,0}\cong \Z\bm{b}^{(1)}_7 \oplus \Z\bm{b}^{(1)}_8$, which are given as $\mu^{(1)}_i = \sum_{j=1}^8 [u^{(1)}]_{ij} n_j$ with 
\begin{align}
u^{(1)}
= 
\begin{array}{c|cccccccc}
&n^\G_1&n^\G_{-1}&n^\G_{(i,-i)}&n^M_1&n^M_{-1}&n^M_{(i,-i)}&n^X_1&n^X_{-1}\\
\hline
\mu^{(1)}_1&  0 & 0 & 1 & 0 & 0 & 0 & 0 & 0 \\
\mu^{(1)}_2&  0 & 1 & 0 & 0 & 0 & 0 & 0 & 0 \\
\mu^{(1)}_3&  0 & 0 & 0 & 0 & 1 & 0 & 0 & 0 \\
\mu^{(1)}_4&  0 & 0 & 0 & 0 & 0 & 1 & 0 & 0 \\
\mu^{(1)}_5&  0 & 0 & 0 & 0 & 0 & 0 & 1 & 0 \\
\mu^{(1)}_6&  0 & 0 & 0 & 1 & 1 & 2 & -1 & 0 \\
\hline 
\mu^{(1)}_7&  1 & 1 & 2 & -1 & -1 & -2 & 0 & 0 \\
\mu^{(1)}_8&  0 & 0 & 0 & -1 & -1 & -2 & 1 & 1 \\
\end{array}. 
\label{eq:c4t_u1}
\end{align}
A nontrivial value of the SIs $(\mu^{(1)}_7,\mu^{(2)}_8) \neq (0,0)$ implies the existence of a gapless point in a 1-cell somewhere. 
Actually, $\mu^{(1)}_7 \neq 0 \ ( \mu^{(2)}_8 \neq 0)$ implies the existence of a fermi line along the 1-cell $\G \to M\ (X \to M)$.

\subsubsection{$E_2^{0,0} \to E_2^{2,-1}$}
In the 2-cell $\alpha$ (shown in Fig.~\ref{fig:c4}), the 2-component gapless Dirac point is protected by the $TC_2(\bk)$ symmetry with the quantized $\pi$-Berry phase, which means $E_1^{2,-1}=\Z_2$. 
Moreover, a  single Dirac point can not be absorbed to 1-cells in a TR and $C_4$ symmetric way, therefore the $\Z_2$ classification persists, $E_2^{2,-1}=\Z_2$. 
Therefore, the second differential $d_2^{0,0}: E_2^{0,0} \to E_2^{2,-1}$ can be nontrivial. 
In fact, one can show that the bases $\bm{b}^{(1)}_2, \bm{b}^{(1)}_3, \bm{b}^{(1)}_6$ create the Dirac point with $\pi$-Berry phase. 
The second differential is given by
\begin{align}
d_2^{0,0}: E_2^{0,0} \to E_2^{2,-1}, \qquad 
d_2^{0,0}(\bm{b}^{(1)}_1,\dots,\bm{b}^{(1)}_6)=\bm{c}^\alpha\ M_{d_2^{0,0}}, 
\end{align}
\begin{align}
M_{d_2^{0,0}}
=
\begin{array}{cccccc|c}
\bm{b}^{(1)}_1&\bm{b}^{(1)}_2&\bm{b}^{(1)}_3&\bm{b}^{(1)}_4&\bm{b}^{(1)}_5&\bm{b}^{(1)}_6\\
\hline
0&1&1&0&0&1&\bm{c}^\alpha\\
\end{array}, 
\end{align}
where we have written the base of $E_2^{2,-1}$ by $\bm{c}^\alpha$. 
This can be verified as follows. 
One can show that in the presence of the $C_4$ symmetry the Berry phase $e^{i \g_{\p \alpha}}$ around the boundary of the 2-cell $\alpha$ shown in Fig.~\ref{fig:c4} can be written as [Chen]
\begin{align}
e^{i \g_{\p \alpha}}
= \frac{w^G_{C_4} w^M_{C_4}}{w^X_{C_2}}, 
\end{align}
where $w^P_g$ is the product of eigenvalues of the symmetry operator $g$ at the high-symmetric point $P$ for the occupied states. 
This reduces to 
\begin{align}
e^{i \g_{\p \alpha}}=(-1)^{n^\G_{-1}+n^M_{-1}+n^X_{-1}}
\label{eq:c4t_berry}
\end{align}
by TRS. 
From Table (\ref{eq:c4t_u1}), we find that if either of the bases $\bm{b}^{(1)}_2,\bm{b}^{(1)}_2, \bm{b}^{(1)}_6$ changes the occupation number, the Berry phase $e^{i \g_{\p \alpha}}$ changes by $-1$.

An alternative brute-force derivation is to classify possible Hamiltonians of the band inversion followed by the creation of a Dirac point for each high-symmetric point. 
We will describe the detail elsewhere. 

According to the strategy in Sec.~\ref{sec:si}, we introduce an integral lift 
\begin{align}
M_{d_2^{0,0}} \mapsto \tilde M_{d_2^{0,0}}
=
\left[
\begin{array}{cccccc|c}
0&1&1&0&0&1&2\\
\end{array}
\right] \in {\rm Mat}_{1 \times 7}(\Z).
\end{align}
The SNF of $\tilde M_{d_2^{0,0}}$ is given by $u' \tilde M_{d_2^{0,0}} v' = 
\left[
\begin{array}{cc}
I_1&O\\
\end{array}
\right]$. 
The SNF of the submatrix $v'_{\rm sub}=\{v'\}_{1\leq i \leq 6,2\leq j \leq 7}$ of $v'$ spanning $\ker d_2^{0,0}$ is given by 
\begin{align}
u^{(2)} v'_{\rm sub} v^{(2)}
=
\left[
\begin{array}{cc}
I_5\\
&2\\
\end{array}
\right], 
\end{align}
from which, we have the group $E_3^{0,0}$ spanned as 
\begin{align}
E_3^{0,0}
=
\bigoplus_{j=1}^5 \Z[\bm{b}^{(2)}_j] \oplus \Z[2\bm{b}^{(2)}_6] 
\end{align}
with $\bm{b}^{(2)}_j=\bm{b}^{(1)}_i[[u^{(2)}]^{-1}]_{ij}$, 
\begin{align}
[u^{(2)}]^{-1}
=
\begin{array}{cccccc|l}
\bm{b}^{(2)}_1&\bm{b}^{(2)}_2&\bm{b}^{(2)}_3&\bm{b}^{(2)}_4&\bm{b}^{(2)}_5&\bm{b}^{(2)}_6&\\
\hline
 1 & 0 & 0 & 0 & 0 & 0 &\bm{b}^{(1)}_1\\
 0 & -1 & 0 & 0 & -1 & -1 &\bm{b}^{(1)}_2\\
 0 & 1 & 0 & 0 & 0 & 0 &\bm{b}^{(1)}_3\\
 0 & 0 & 1 & 0 & 0 & 0 &\bm{b}^{(1)}_4\\
 0 & 0 & 0 & 1 & 0 & 0 &\bm{b}^{(1)}_5\\
 0 & 0 & 0 & 0 & 1 & 0 &\bm{b}^{(1)}_6\\
\end{array}. 
\end{align}
The SI for detecting $E_2^{0,0}/E_3^{0,0} = \Z_2[\bm{b}^{(2)}_6]$ is given by $\mu^{(2)}_i = \sum_{j=1}^6 [u^{(2)}]_{ij} \mu^{(1)}_j$, 
\begin{align}
u^{(2)}
=
\begin{array}{c|cccccc}
&\mu^{(1)}_1&\mu^{(1)}_2&\mu^{(1)}_3&\mu^{(1)}_4&\mu^{(1)}_5&\mu^{(1)}_6\\
\hline
\mu^{(2)}_1& 1 & 0 & 0 & 0 & 0 & 0 \\
\mu^{(2)}_2& 0 & 0 & 1 & 0 & 0 & 0 \\
\mu^{(2)}_3& 0 & 0 & 0 & 1 & 0 & 0 \\
\mu^{(2)}_4& 0 & 0 & 0 & 0 & 1 & 0 \\
\mu^{(2)}_5& 0 & 0 & 0 & 0 & 0 & 1 \\
 \hline
\mu^{(2)}_6& 0 & -1 & -1 & 0 & 0 & -1 \\
\end{array}. 
\end{align} 
We have the $\Z_2$-valued SI 
\begin{align}
\mu^{(2)}_6
=
-\mu^{(1)}_2-\mu^{(1)}_3-\mu^{(1)}_6 
=-n^G_{-1}-n^M_{-1}-(n^M_1+n^M_{-1}+2n^M_{(i,-i)}-n^X_1) \quad {\rm mod\ 2}.
\end{align}
Note that modulo $\im d_1^{0,0}$, namely by using $\mu^{(1)}_8=-n^M_1-n^M_{-1}-2n^M_{(i,-i)}+n^X_1+n^X_{-1}=0$, this is recast as (\ref{eq:c4t_berry}). 
Provided that the first SIs are trivial $\mu^{(1)}_7=\mu^{(1)}_8=0$, $\mu^{(2)}_6 \neq 0$ implies the existence of a Dirac point with the quantized $\pi$-Berry phase in the 2-cell $\alpha$.

\subsubsection{$f^{\rm AI}_{d\leq 2}(\{{\rm AI}_{d \leq 2}\}) \cong E^3_{0,0}$}
Since no building-block TIs in 1- and 2-dimensions exist for class AI systems, any elements of $E_3^{0,0}$ should be represented as an abelian sum of atomic insulators.
Let us confirm the equivalence $\im f^{\rm AI}_{d \leq 2} \cong E^3_{0,0}$ directly. 
The group $\{{\rm AI}_{d \leq 2}\}$ is generated by atomic insulators in the unit cell. 
Let us write the atomic insulator induced by the irrep $\rho$ at the coordinate $\bm{x}$ in the unit cell by $\bm{a}^{\bm{x}}_\rho$. 
There are eight generators with the data of topological invariants in the $k$-space, 
\begin{align}
\begin{array}{c|ccc|ccc|cc}
&n^\G_1&n^\G_{-1}&n^\G_{(i,-i)}&n^M_1&n^M_{-1}&n^M_{(i,-i)}&n^X_1&n^X_{-1}\\
\hline 
\bm{a}^{(0,0)}_1&1&0&0&1&0&0&1&0 \\
\bm{a}^{(0,0)}_{-1}&0&1&0&0&1&0&1&0\\
\bm{a}^{(0,0)}_{(i,-i)}&0&0&1&0&0&1&0&2 \\
\bm{a}^{(\frac{1}{2},\frac{1}{2})}_1&1&0&0&0&1&0&0&1 \\
\bm{a}^{(\frac{1}{2},\frac{1}{2})}_{-1}&0&1&0&1&0&0&0&1\\
\bm{a}^{(\frac{1}{2},\frac{1}{2})}_{(i,-i)}&0&0&1&0&0&1&2&0 \\
\bm{a}^{(\frac{1}{2},0)}_1&1&1&0&0&0&1&1&1 \\
\bm{a}^{(\frac{1}{2},0)}_{-1}&0&0&1&1&1&0&1&1 \\
\end{array}
=: (M_{f_{d \leq 2}})^T.
\label{tab:c4_top_inv_AI}
\end{align}
This gives the homomorphism $f^{\rm AI}_{d \leq 2}: \{{\rm AI}_{d\leq 2}\} \to E_1^{0,0}$ and that in the basis of $E_3^{0,0}$ as in 
\begin{align}
f^{\rm AI}_{d\leq 2}
(\bm{a}^{(0,0)}_1,\dots,\bm{a}^{(\frac{1}{2},0)}_{-1})
&=
(\bm{b}^\G_1, \dots, \bm{b}^X_{-1}) M_{f^{\rm AI}_{d\leq 2}} \nonumber \\
&=
(\bm{b}^{(2)}_1, \dots, \bm{b}^{(2)}_5,2\bm{b}^{(2)}_6,\bm{b}^{(1)}_7,\bm{b}^{(1)}_8) 
\left[
\begin{array}{c|c}
\left[
\begin{array}{cc}
I_5\\
&\frac{1}{2}\\
\end{array}
\right]
u^{(2)}\\
\hline
&I_2\\
\end{array}
\right]
u^{(1)} 
M_{f^{\rm AI}_{d\leq 2}}, 
\end{align}
where 
\begin{align}
\left[
\begin{array}{c|c}
\left[
\begin{array}{cc}
I_5\\
&\frac{1}{2}\\
\end{array}
\right]
u^{(2)}\\
\hline
&I_2\\
\end{array}
\right]
u^{(1)} 
M_{f^{\rm AI}_{d\leq 2}}
= 
\left[
\begin{array}{cccccccc}
 0 & 0 & 1 & 0 & 0 & 1 & 0 & 1 \\
 0 & 1 & 0 & 1 & 0 & 0 & 0 & 1 \\
 0 & 0 & 1 & 0 & 0 & 1 & 1 & 0 \\
 1 & 1 & 0 & 0 & 0 & 2 & 1 & 1 \\
 0 & 0 & 2 & 1 & 1 & 0 & 1 & 1 \\
 0 & -1 & -1 & -1 & -1 & 0 & -1 & -1 \\
\hline
 0 & 0 & 0 & 0 & 0 & 0 & 0 & 0 \\
 0 & 0 & 0 & 0 & 0 & 0 & 0 & 0 \\ 
\end{array}
\right]
=:
\left[
\begin{array}{c}
\tilde M^{\rm sub}_{f^{\rm AI}_{d \leq 2}}\\
O\\
\end{array}
\right]. 
\end{align}
This has the SNF $\tilde M^{\rm sub}_{f^{\rm AI}_{d \leq 2}}
\sim 
\left[
\begin{array}{cc}
I_6&O\\
\end{array}
\right]$, which means $\im f^{\rm AI}_{d \leq 2} \cong E_3^{0,0}$.

\subsubsection{$\{{\rm AI}_{d \leq 1}\} \to E_1^{0,0}$}
\label{sec:2d_c4_ai1}
A nontrivial inclusion of insulators starts at the atomic insulators not located at the corner of the unit cell, 
\begin{align}
f^{\rm AI}_{d \leq 1}: \{{\rm AI}_{d \leq 1}\} \to E^1_{0,0}. 
\end{align}
The group $\{{\rm AI}_{d \leq 1}\}$ is generated by atomic insulators at high-symmetric points $(0,0), (\frac{1}{2},0), (0,\frac{1}{2})$ in the unit cell. 
We have five generators with the data of topological invariants in the $k$-space, 
\begin{align}
\begin{array}{c|ccc|ccc|cc}
&n^\G_1&n^\G_{-1}&n^\G_{(i,-i)}&n^M_1&n^M_{-1}&n^M_{(i,-i)}&n^X_1&n^X_{-1}\\
\hline 
\bm{a}^{(0,0)}_1&1&0&0&1&0&0&1&0 \\
\bm{a}^{(0,0)}_{-1}&0&1&0&0&1&0&1&0\\
\bm{a}^{(0,0)}_{(i,-i)}&0&0&1&0&0&1&0&2 \\
\bm{a}^{(\frac{1}{2},0)}_1&1&1&0&0&0&1&1&1 \\
\bm{a}^{(\frac{1}{2},0)}_{-1}&0&0&1&1&1&0&1&1 \\
\end{array}
=: (M_{f_{d \leq 1}})^T.
\end{align}
This gives the homomorphism in the basis of $E_3^{0,0}$. 
\begin{align}
f^{\rm AI}_{d\leq 1}(\bm{a}^{(0,0)}_1,\dots,\bm{a}^{(\frac{1}{2},0)}_{-1})
&=
(\bm{b}^\G_1, \dots, \bm{b}^X_{-1}) M_{f^{\rm AI}_{d\leq 1}} \nonumber \\
&=
(\bm{b}^{(2)}_1, \dots, \bm{b}^{(2)}_5,2\bm{b}^{(2)}_5,\bm{b}^{(1)}_7,\bm{b}^{(1)}_8) 
\left[
\begin{array}{c|c}
\left[
\begin{array}{cc}
I_5\\
&\frac{1}{2}\\
\end{array}
\right]
u^{(2)}\\
\hline
&I_2\\
\end{array}
\right]
u^{(1)} M_{f^{\rm AI}_{d\leq 1}}, 
\end{align}
where 
\begin{align}
\left[
\begin{array}{c|c}
\left[
\begin{array}{cc}
I_5\\
&\frac{1}{2}\\
\end{array}
\right]
u^{(2)}\\
\hline
&I_2\\
\end{array}
\right]
u^{(1)} M_{f^{\rm AI}_{d\leq 1}}
=
\left[
\begin{array}{ccccc}
 0 & 0 & 1 & 0 & 1 \\
 0 & 1 & 0 & 0 & 1 \\
 0 & 0 & 1 & 1 & 0 \\
 1 & 1 & 0 & 1 & 1 \\
 0 & 0 & 2 & 1 & 1 \\
 0 & -1 & -1 & -1 & -1 \\
\hline
0&0&0&0&0\\
 0&0&0&0&0\\
\end{array}
\right]
=: 
\left[
\begin{array}{c}
\tilde M^{\rm sub}_{f^{\rm AI}_{d \leq 1}}\\
O
\end{array}
\right].
\end{align}
The SNF of this is given by 
\begin{align}
u^{(3)} \tilde M^{\rm sub}_{f^{\rm AI}_{d \leq 1}} v^{(3)}
=
\left[
\begin{array}{c|c}
I_4&O\\
\hline
O&O
\end{array}
\right], 
\end{align}
so we introduce the new basis $(\bm{b}^{(3)}_1,\dots,\bm{b}^{(3)}_6)
=
(\bm{b}^{(2)}_1,\dots,\bm{b}^{(2)}_5,2\bm{b}^{(2)}_6) [u^{(3)}]^{-1}$ with 
\begin{align}
[u^{(3)}]^{-1} 
=
\begin{array}{cccccc|l}
\bm{b}^{(3)}_1&\bm{b}^{(3)}_2&\bm{b}^{(3)}_3&\bm{b}^{(3)}_4&\bm{b}^{(3)}_5&\bm{b}^{(3)}_6&\\
\hline
 0 & 0 & 1 & 0 & 0 & 0 &\bm{b}^{(2)}_1\\
 0 & 1 & 0 & 0 & 0 & 0 &\bm{b}^{(2)}_2\\
 0 & 0 & 1 & 1 & 0 & 0 &\bm{b}^{(2)}_3\\
 1 & 1 & 0 & 1 & 0 & 0 &\bm{b}^{(2)}_4\\
 0 & 0 & 2 & 1 & 1 & 0 &\bm{b}^{(2)}_5\\
 0 & -1 & -1 & -1 & 0 & 1 &2\bm{b}^{(2)}_6\\
\end{array}. 
\end{align}
We find that the image of $f^{\rm AI}_{d\leq 1} \subset E_3^{0,0}$ is spanned as $\im f^{\rm AI}_{d\leq 1} = \bigoplus_{j=1}^4 \Z[\bm{b}^{(3)}_j]$. 
The SIs detecting the quotient group $E_3^{0,0}/\im f^{\rm AI}_{d\leq 1}=\Z[\bm{b}^{(3)}_5]\oplus\Z[\bm{b}^{(3)}_6]$ are given by 
$\xi^{(1)}_i=\sum_{j=1}^6 [u^{(3)}]_{ij} \mu^{(2)}_j$ with 
\begin{align}
u^{(3)}
=
\begin{array}{c|cccccc}
&\mu^{(2)}_1&\mu^{(2)}_2&\mu^{(2)}_3&\mu^{(2)}_4&\mu^{(2)}_5&\mu^{(2)}_6\\
\hline
\xi^{(1)}_1& 1 & -1 & -1 & 1 & 0 & 0 \\
\xi^{(1)}_2& 0 & 1 & 0 & 0 & 0 & 0 \\
\xi^{(1)}_3& 1 & 0 & 0 & 0 & 0 & 0 \\
\xi^{(1)}_4& -1 & 0 & 1 & 0 & 0 & 0 \\
 \hline
\xi^{(1)}_5& -1 & 0 & -1 & 0 & 1 & 0 \\
\xi^{(1)}_6& 0 & 1 & 1 & 0 & 0 & 1 \\
\end{array}. 
\end{align} 
Provided that the lower SIs $\mu^{(1)}_7, \mu^{(1)}_8, \mu^{(2)}_6$ are trivial, $(\xi^{(1)}_5,\xi^{(1)}_6) \neq (0,0)$ implies the existence of a corner state. 
See Sec.~\ref{sec:c4t_ex} for demonstration.

\subsubsection{$\{{\rm AI}_{d \leq 0}\} \to E^1_{0,0}$}
\label{sec:2d_c4_ai0}
The last step is to evaluate the SI for an ingap edge state. 
Let $\{{\rm AI}_{d \leq 0}\}$ be the abelian group generated by atomic insulators at $(0,0)$ in the unit cell, says, $\{ \bm{a}^{(0,0)}_1,\bm{a}^{(0,0)}_{-1},\bm{a}^{(0,0)}_{(i,-i)}\}$. 
The homomorphism 
\begin{align}
f^{\rm AI}_{d\leq 0}(\bm{a}^{(0,0)}_1,\bm{a}^{(0,0)}_{-1},\bm{a}^{(0,0)}_{(i,-i)})
=
(\bm{b}^\G_1, \dots, \bm{b}^X_{-1}) M_{f^{\rm AI}_{d\leq 0}} 
\end{align}
is given by 
\begin{align}
\begin{array}{c|ccc|ccc|cc}
&n^\G_1&n^\G_{-1}&n^\G_{(i,-i)}&n^M_1&n^M_{-1}&n^M_{(i,-i)}&n^X_1&n^X_{-1}\\
\hline 
\bm{a}^{(0,0)}_1&1&0&0&1&0&0&1&0 \\
\bm{a}^{(0,0)}_{-1}&0&1&0&0&1&0&1&0\\
\bm{a}^{(0,0)}_{(i,-i)}&0&0&1&0&0&1&0&2 \\
\end{array}
=: (M_{f_{d \leq 0}})^T.
\end{align}
To compute the SI detecting edge states, we rewrite the homomorphism $f^{\rm AI}_{d\leq 0}$ in the basis of $\im f^{\rm AI}_{d\leq 1}$ as in 
\begin{align}
f^{\rm AI}_{d\leq 0}(\bm{a}^{(0,0)}_1,\bm{a}^{(0,0)}_{-1},\bm{a}^{(0,0)}_{(i,-i)})
=
(\bm{b}^{(3)}_1, \dots,\bm{b}^{(3)}_6,\bm{b}^{(1)}_7,\bm{b}^{(1)}_8) 
\left[
\begin{array}{c|c}
u^{(3)}\left[
\begin{array}{cc}
I_5\\
&\frac{1}{2}\\
\end{array}
\right]
u^{(2)}\\
\hline
&I_2\\
\end{array}
\right]
u^{(1)} M_{f^{\rm AI}_{d\leq 0}}, 
\end{align}
where 
\begin{align}
\left[
\begin{array}{c|c}
u^{(3)}\left[
\begin{array}{cc}
I_5\\
&\frac{1}{2}\\
\end{array}
\right]
u^{(2)}\\
\hline
&I_2\\
\end{array}
\right]
u^{(1)} M_{f^{\rm AI}_{d\leq 0}}
=
\left[
\begin{array}{c}
I_3\\
O\\
\end{array}
\right].
\end{align}
Therefore, the basis $\{\bm{b}^{(3)}_j\}_{j=1}^3$ is already the basis for $\im f^{\rm AI}_{d\leq 0}$. 
The SIs detecting the quotient group $\im f^{\rm AI}_{d\leq 1}/\im f^{\rm AI}_{d\leq 0}=\Z[\bm{b}^{(3)}_4]$ is given by $\xi^{(2)}_4=\xi^{(1)}_4$. 
Provided that the SIs $\mu^{(1)}_7, \mu^{(1)}_8, \mu^{(2)}_6,\xi^{(1)}_6,\xi^{(1)}_5$ are trivial, the nontrivial SI $\xi^{(1)}_4 \neq 0$ implies the existence of an edge state without corner states. 

Table~\ref{tab:si_c4t} summarizes the SIs we got in this section. 

\begin{table}
\centering
\caption{SIs for spinless electrons with TR and $C_4$ symmetry in $2d$}
\label{tab:si_c4t}
\begin{align*}
\begin{array}{lcccccccccl}
\mbox{SI}&\mbox{Range}&n^\G_1&n^\G_{-1}&n^\G_{(i,-i)}&n^M_1&n^M_{-1}&n^M_{(i,-i)}&n^X_1&n^X_{-1}&\mbox{Target}\\
\hline
\xi^{(2)}_4&\Z& 0 & 0 & -1 & 0 & 0 & 1 & 0 & 0 & \mbox{Ingap edge states}\\
\hline
\xi^{(1)}_5&\Z& 0 & 0 & -1 & 1 & 1 & 1 & -1 & 0 & \mbox{Ingap corner states} \\
\xi^{(1)}_6&\Z& 0 & -\frac{1}{2} & 0 & -\frac{1}{2} & 0 & 0 & \frac{1}{2} & 0 &\\
\hline
\mu^{(2)}_6&\Z/2\Z& 0 & -1 & 0 & -1 & -2 & -2 & 1 & 0 & \mbox{Gapless states in 2-cell}\\
\hline
\mu^{(1)}_7&\Z& 1 & 1 & 2 & -1 & -1 & -2 & 0 & 0 & \mbox{Gapless states in 1-cell}\\
\mu^{(1)}_8&\Z& 0 & 0 & 0 & -1 & -1 & -2 & 1 & 1 \\
\end{array}
\end{align*}
\end{table}

\subsubsection{Some models}
\label{sec:c4t_ex}
In this section, we demonstrate how the SIs gapless phases and ingap boundary states work. 

The first example is the spinless Hamiltonian on the square lattice with a nearest neighbor hoping.  
\begin{align}
&H_1(\bk)
=
-\cos k_x - \cos k_y - \mu, \qquad 
C_4(\bk)=1, \qquad T=K. 
\end{align}
When the chemical potential is set in $-2 <\mu<0$, a fermi line exists enclosing the $\G$ point. 

The second example is the following tight-binding model with four dof sitting the center of the unit cell.~\cite{ono2019difficulties,BenalcazarCorner}
We consider the $\pi$-flux loop hopping around the four corners of the unit cell. 
The Hamiltonian in the $k$-space is 
\begin{align}
&H_2(\bk)
=
-(\cos k_x \s_x + \sin k_x \s_y \tau_z+\cos k_y \s_x \tau_x + \sin k_y \s_x \tau_y), \qquad 
C_4(\bk)=\begin{pmatrix}
0&0&0&1\\
1&0&0&0\\
0&1&0&0\\
0&0&1&0\\
\end{pmatrix}, \qquad T=K. 
\end{align}
This has four flat bands with Bloch energies $E_{\bk}=2,0,0,-2$, of which the Wanner orbitals are located at the corner of the unit cell. 
Here, we focus on the lowest energy band $E_{\bk}=-2$ that has the Bloch state $\ket{u(\bk)}=\frac{1}{2}(1,e^{-ik_x},e^{-ik_x-ik_y},e^{-ik_y})$.

The third example is a semimal phase. 
Consider the following 2 by 2 Hamiltonian 
\begin{align}
&H_3(\bk)
=(\cos k_x-\cos k_y)\s_x+(\cos k_x+\cos k_y)\s_z, \qquad 
C_4(\bk)=\s_z, \qquad T=K.
\end{align}
This model has a Dirac point with the $\pi$-Berry phase in the quarter of the BZ. 

The fourth example is given by stacking two layers of the semimetal phase $H_3(\bk)$.
One can induce a finite mass gap to the Dirac points as 
\begin{align}
&H_4(\bk)
=(\cos k_x-\cos k_y)\s_x+(\cos k_x+\cos k_y)\s_z+\epsilon \sin k_x \sin k_y \s_y \tau_y, \qquad 
C_4(\bk)=\s_z, \qquad 
T=K, 
\end{align}
with $\epsilon$ a small constant. 
The occupied states of $H_4(\bk)$ is the direct sum of Chern insulators with $C=2$ and $C=-2$, and is a fragile topological phase since it cannot be represented as a linear combination of atomic insulators listed in Table (\ref{tab:c4_top_inv_AI}). 
Also, we see the occupied states has a corner state.~\cite{BenalcazarCorner}

The values of the SIs for models $H_1,\dots,H_4$ are listed below. 
\begin{align}
\begin{array}{l|cccccccc|cccccc}
\mbox{Model}&n^\G_1&n^\G_{-1}&n^\G_{(i,-i)}&n^M_1&n^M_{-1}&n^M_{(i,-i)}&n^X_1&n^X_{-1}&
\xi^{(2)}_4&\xi^{(1)}_5&\xi^{(1)}_6&\mu^{(2)}_6&\mu^{(1)}_7&\mu^{(1)}_8\\
\hline
H_1&1&0&0&0&0&0&0&0&0&0&0&0&1&0\\
H_2&1&0&0&0&1&0&0&1&0&1&0&0&0&0\\
H_3&0&1&0&1&0&0&1&0&0&0&-\frac{1}{2}&1&0&0\\
H_4&0&2&0&2&0&0&2&0&0&0&-1&0&0&0\\
\end{array}
\end{align}
We see that the SIs correctly capture gapless states and ingap boundary states.

\subsection{$3d$ TR-symmetric odd-parity SCs}
The last example is TR-symmetric odd-parity SCs in the 3-dimensional cubic lattice. 
The symmetry constraints are summarized as 
\begin{align}
&TH(\bk)T^{-1}=H(-\bk), \qquad T^2=-1, \nonumber \\
&CH(\bk)C^{-1}=-H(-\bk), \qquad C^2=1, \nonumber \\
&P(\bk)H(\bk)P(\bk)^{-1}=H(-\bk), \qquad P(-\bk)P(\bk)=1, \nonumber \\
&TP(\bk)=P(-\bk)T, \qquad CP(\bk)=-P(-\bk)C.
\end{align}
At eight high-symmetric points in the BZ, the effective AZ class is class AII, so we have $E_1^{0,0}=\Z^8$ generated respectively by 
\begin{align}
\bm{b}^k
=
\left[
E=\C^4,
T=is_yK, C=\tau_xK, P=\tau_z, H=-\tau_z, H_0=\tau_z
\right] 
\end{align}
characterized by the $\Z$ invariant 
\begin{align}
n^k=\frac{1}{2}\big\{N_+[H(k)]-N_+[H_0(k)]\big\}  \in \Z
\label{eq:3dtsc_top_inv}
\end{align}
at eight high-symmetric points. 
Here, $N_\pm[H]$ denotes the number of occupied states of $H$ with the positive/negative eigenstates of $P(k)$, and the prefactor is due to the Kramaers degeneracy.
In the weak coupling limit, $n^k$ is simplified to the difference of the numbers of positive and negative parity occupied states 
\begin{align}
n^k
=
\frac{1}{2}\big\{N_+[h(k)]-N_-[h(k)]\big\} \in \Z.
\label{eq:3dtsc_top_inv_wcl}
\end{align}

At generic points in the BZ, the EAZ is class CII, meaning no stable gapless points in 1,2, and 3-cells. 
Thus, we have $E_4^{0,0}=E_3^{0,0}=E_2^{0,0}=E_1^{0,0}$. 
Every element in $E_1^{0,0}$ can be represented by a fully gapped superconductor.

\subsubsection{Building-block states}
To compute the homomorphisms $f^{\rm TI}_{d\leq p}: \{{\rm TI}_{d\leq p} \to E_1^{0,0}, (p=0,1,2)$, we should list all possible SCs with space dimension less than $p$. 
For inversion symmetry, the layer construction suffices.~\footnote{This is not always the case.
In general, a generating model forms a pattern of local building-block states in the real space that may not be defined on a layer.  
}
Let us introduce building-block $0d$, $1d$, $2d$, and $3d$ states as 
\begin{align}
&H_{0d}
= -\tau_z, \\
&H_{1d,i}(k_i)
= (-\cos k_i - \mu) \tau_z + \sin k_i s_z \tau_x, \qquad (-1<\mu<1), \\
&H_{2d,ij}(k_i,k_j)
= (-\cos k_i -\cos k_j - \mu) \tau_z + \sin k_i s_z \tau_x + \sin k_j \tau_y, \qquad (-2 < \mu < 0), \\
&H_{3d}(k_x,k_y,k_z)
= (-\cos k_x -\cos k_y -\cos k_z-\mu) \tau_z + \sin k_x s_z \tau_x + \sin k_y \tau_y+\sin k_z s_x\tau_x, \qquad (-3<\mu<-1), 
\label{eq:3d_tsc_model}
\end{align}
with TRS and PHS operators $T=is_yK, C=\tau_xK$.
The group $\{{\rm TI}_{d \leq p}\}$ is generated by the triples of these building-block states located at either of eight inversion centers relative to the vacuum Hamiltonian $H_0=\tau_z$. 
The $0d$ atomic insulator $H_{0d}$ can be put on either of eight high-symmetric points in the unit cell. 
The triples $[E,H,H_0]$ have the following topological invariants defined by (\ref{eq:3dtsc_top_inv}).
\begin{align}
[M_{f^{\rm TI}_{d=0}}]^T=
\begin{array}{ccc|cccccccc}
&P(\bk)&H&n^{000}&n^{100}&n^{010}&n^{001}&n^{110}&n^{011}&n^{101}&n^{111}\\
\hline
\bm{a}_{0d}^{(0,0,0)}&\tau_z&H_{0d}&
1&1&1&1&1&1&1&1\\
\bm{a}_{0d}^{(\frac{1}{2},0,0)}&e^{-ik_x}\tau_z&H_{0d}&
1&-1&1&1&-1&1&-1&-1\\
\bm{a}_{0d}^{(0,\frac{1}{2},0)}&e^{-ik_y}\tau_z&H_{0d}&
1&1&-1&1&-1&-1&1&-1\\
\bm{a}_{0d}^{(0,0,\frac{1}{2})}&e^{-ik_z}\tau_z&H_{0d}&
1&1&1&-1&1&-1&-1&-1\\
\bm{a}_{0d}^{(\frac{1}{2},\frac{1}{2},0)}&e^{-i(k_x+k_y)}\tau_z&H_{0d}&
1&-1&-1&1&1&-1&-1&1\\
\bm{a}_{0d}^{(0,\frac{1}{2},\frac{1}{2})}&e^{-i(k_y+k_z)}\tau_z&H_{0d}&
1&1&-1&-1&-1&1&-1&1\\
\bm{a}_{0d}^{(\frac{1}{2},0,\frac{1}{2})}&e^{-i(k_z+k_x)}\tau_z&H_{0d}&
1&-1&1&-1&-1&-1&1&1\\
\bm{a}_{0d}^{(\frac{1}{2},\frac{1}{2},\frac{1}{2})}&e^{-i(k_x+k_y+k_z)}\tau_z&H_{0d}&
1&-1&-1&-1&1&1&1&-1\\
\end{array}. 
\end{align}
Here the superscript of $\bm{a}^{\bm{x}}$ represents the Wyckoff position on which the Hamiltonian is defined, and that of $n^{lmn}$ represents the high-symmetric points $\bk = \pi(l,m,n)$. 
Similarly, $1d$ building-blocks are listed as~\footnote{
Here, it is not needed to add the model $H_{1d,x}$ on the Wyckoff position $(\frac{1}{2},0,0)$, since the triple $[E,H_{1d,x},H_0]$ of it has the same topological invariants as that on the Wyckoff position $(0,0,0)$. 
}
\begin{align}
[M_{f^{\rm TI}_{d=1}}]^T=
\begin{array}{ccc|ccccccccc}
&P(\bk)&H&n^{000}&n^{100}&n^{010}&n^{001}&n^{110}&n^{011}&n^{101}&n^{111}\\
\hline
\bm{a}_{1d,x}^{(0,0,0)}&\tau_z&H_{1d,x}&
1&0&1&1&0&1&0&0\\
\bm{a}_{1d,x}^{(0,\frac{1}{2},0)}&e^{-ik_y}\tau_z&H_{1d,x}&
1&0&-1&1&0&-1&0&0\\
\bm{a}_{1d,x}^{(0,0,\frac{1}{2})}&e^{-ik_z}\tau_z&H_{1d,x}&
1&0&1&-1&0&-1&0&0\\
\bm{a}_{1d,x}^{(0,\frac{1}{2},\frac{1}{2})}&e^{-i(k_y+k_z)}\tau_z&H_{1d,x}&
1&0&-1&-1&0&1&0&0\\
\hline
\bm{a}_{1d,y}^{(0,0,0)}&\tau_z&H_{1d,y}&
1&1&0&1&0&0&1&0\\
\bm{a}_{1d,y}^{(\frac{1}{2},0,0)}&e^{-ik_x}\tau_z&H_{1d,y}&
1&-1&0&1&0&0&-1&0\\
\bm{a}_{1d,y}^{(0,0,\frac{1}{2})}&e^{-ik_z}\tau_z&H_{1d,y}&
1&1&0&-1&0&0&-1&0\\
\bm{a}_{1d,y}^{(\frac{1}{2},0,\frac{1}{2})}&e^{-i(k_x+k_z)}\tau_z&H_{1d,y}&
1&-1&0&-1&0&0&1&0\\
\hline
\bm{a}_{1d,z}^{(0,0,0)}&\tau_z&H_{1d,z}&
1&1&1&0&1&0&0&0\\
\bm{a}_{1d,z}^{(\frac{1}{2},0,0)}&e^{-ik_x}\tau_z&H_{1d,z}&
1&-1&1&0&-1&0&0&0\\
\bm{a}_{1d,z}^{(0,\frac{1}{2},0)}&e^{-ik_y}\tau_z&H_{1d,z}&
1&1&-1&0&-1&0&0&0\\
\bm{a}_{1d,z}^{(\frac{1}{2},\frac{1}{2},0)}&e^{-i(k_x+k_y)}\tau_z&H_{1d,z}&
1&-1&-1&0&1&0&0&0\\
\end{array}. 
\end{align}
Similarly, we have $2d$ building blocks 
\begin{align}
[M_{f^{\rm TI}_{d=2}}]^T=
\begin{array}{ccc|ccccccccc}
&P(\bk)&H&n^{000}&n^{100}&n^{010}&n^{001}&n^{110}&n^{011}&n^{101}&n^{111}\\
\hline
\bm{a}_{2d,xy}^{(0,0,0)}&\tau_z&H_{2d,xy}&
1&0&0&1&0&0&0&0\\
\bm{a}_{2d,xy}^{(0,0,\frac{1}{2})}&e^{-ik_z}\tau_z&H_{2d,xy}&
1&0&0&-1&0&0&0&0\\
\hline
\bm{a}_{2d,yz}^{(0,0,0)}&\tau_z&H_{2d,yz}&
1&1&0&0&0&0&0&0\\
\bm{a}_{2d,yz}^{(\frac{1}{2},0,0)}&e^{-ik_x}\tau_z&H_{2d,yz}&
1&-1&0&0&0&0&0&0\\
\hline
\bm{a}_{2d,xz}^{(0,0,0)}&\tau_z&H_{2d,xz}&
1&0&1&0&0&0&0&0\\
\bm{a}_{2d,xz}^{(0,\frac{1}{2},0)}&e^{-ik_y}\tau_z&H_{2d,xz}&
1&0&-1&0&0&0&0&0\\
\end{array}.
\end{align}
Finally, the $3d$ building block has the data 
\begin{align}
[M_{f^{\rm TI}_{d=3}}]^T=
\begin{array}{ccc|ccccccccc}
&P(\bk)&H&n^{000}&n^{100}&n^{010}&n^{001}&n^{110}&n^{011}&n^{101}&n^{111}\\
\hline
\bm{a}_{3d}^{(0,0,0)}&\tau_z&H_{3d}&
1&0&0&0&0&0&0&0\\
\end{array}.
\end{align}

\subsubsection{$\{{\rm TI}_{d \leq 3}\} \to E_1^{0,0}$}
Let us write the basis of $E_1^{0,0}$ by $(\bm{b}^{000},\dots,\bm{b}^{111})$. 
Every element of $E_1^{0,0}(=E_4^{0,0})$ should be represented as a fully gapped SC. 
To see this, consider the homomorohism $f^{\rm TI}_{d \leq 3}: \{{\rm TI}_{d\leq 3}\} \to E_1^{0,0}$, 
\begin{align}
&f^{\rm TI}_{d \leq 3}(\bm{a}_{1d}^{0,0,0},\dots,\bm{a}_{3d}^{(0,0,0)})
=
(\bm{b}^{000},\dots,\bm{b}^{111}) M_{f^{\rm TI}_{d \leq 3}}, 
\end{align}
where 
\begin{align}
M_{f^{\rm TI}_{d \leq 3}}
=
\left[
\begin{array}{cccc}
M_{f^{\rm TI}_{d=0}}&
M_{f^{\rm TI}_{d=1}}&
M_{f^{\rm TI}_{d=2}}&
M_{f^{\rm TI}_{d=3}}
\end{array}
\right].
\end{align}
The SNF of $M_{f^{\rm TI}_{d \leq 3}}$ is given by 
\begin{align}
M_{f^{\rm TI}_{d \leq 3}}
\sim 
\left[
\begin{array}{cc}
I_8&O\\
\end{array}
\right].
\end{align}
This means $\im f^{\rm TI}_{d\leq 3} \cong E_1^{0,0}$.

\subsubsection{SIs for TSCs}
A nontrivial subgroup starts from the homomorphism $f^{\rm TI}_{d \leq 2}: \{{\rm TI}_{d \leq 2}\} \to E_1^{0,0}$, 
\begin{align}
&f^{\rm TI}_{d \leq 2}(\bm{a}_{1d}^{(0,0,0)},\dots,\bm{a}_{2d,xz}^{(0,\frac{1}{2},0)})
=
(\bm{b}^{000},\dots,\bm{b}^{111}) M_{f^{\rm TI}_{d \leq 2}} 
\end{align}
with 
\begin{align}
M_{f^{\rm TI}_{d \leq 2}}
=
\left[
\begin{array}{cccc}
M_{f^{\rm TI}_{d=0}}&
M_{f^{\rm TI}_{d=1}}&
M_{f^{\rm TI}_{d=2}}
\end{array}
\right].
\end{align}
The SNF of $M_{f^{\rm TI}_{d \leq 2}}$ is given by 
\begin{align}
u^{(1)} M_{f^{\rm TI}_{d \leq 2}} v^{(1)}
=
\left[
\begin{array}{cc|c}
I_7&&O\\
&2&O\\
\end{array}
\right].
\end{align}
Thus, we have a $\Z_2$-valued SI $\nu^{(1)}_8$ detecting the 1st-order TSC $E_1^{0,0}/\im f^{\rm TI}_{d \leq 2} = \im f^{\rm TI}_{d\leq 3}/\im f^{\rm TI}_{d \leq 2} = \Z_2$. 
We summarize the explicit formulas of SIs later. 

Next, we shall compute the homomorphism $\{{\rm TI}_{d \leq 1}\} \to E_1^{0,0}$, 
\begin{align}
f^{\rm TI}_{d\leq 1}(\bm{a}_{1d}^{0,0,0},\dots,\bm{a}_{1d,z}^{(\frac{1}{2},\frac{1}{2},0)})
=
(\bm{b}^{000},\dots,\bm{b}^{111}) M_{f^{\rm TI}_{d \leq 1}} 
\end{align}
with 
\begin{align}
M_{f^{\rm TI}_{d \leq 1}}
=
\left[
\begin{array}{cccc}
M_{f^{\rm TI}_{d=0}}&
M_{f^{\rm TI}_{d=1}}
\end{array}
\right].
\end{align}
The SNF of the homomorphism $f^{\rm TI}_{d\leq 1}$ in the basis of $\im f^{\rm TI}_{d \leq 2}$ is given by 
\begin{align}
u^{(2)} 
\left[
\begin{array}{cc}
I_7\\
&\frac{1}{2}\\
\end{array}
\right]
u^{(1)} M_{f^{\rm TI}_{d \leq 1}} v^{(2)}
=
\left[
\begin{array}{cc|c}
I_4&\\
&2I_4&O\\
\end{array}
\right].
\end{align}
Thus, we have four $\Z_2$-valued SIs $\nu^{(2)}_5,\dots,\nu^{(2)}_8$ detecting 2nd-order TSCs. 

At last, we compute the homomorphism $\{{\rm TI}_{d=0}\} \to E_1^{0,0}$, 
\begin{align}
f^{\rm TI}_{d=1}(\bm{a}_{1d}^{(0,0,0)},\dots,\bm{a}_{0d}^{(\frac{1}{2},\frac{1}{2},\frac{1}{2})})
=
(\bm{b}^{000},\dots,\bm{b}^{111}) M_{f^{\rm TI}_{d=0}}.
\end{align}
The SNF of the homomorphism $f^{\rm TI}_{d=0}$ in the basis of $\im f^{\rm TI}_{d \leq 1}$ is given by 
\begin{align}
u^{(3)}
\left[
\begin{array}{cc}
I_4\\
&\frac{1}{2}I_4\\
\end{array}
\right]
u^{(2)} 
\left[
\begin{array}{cc}
I_7\\
&\frac{1}{2}\\
\end{array}
\right]
u^{(1)} M_{f^{\rm TI}_{d=0}} v^{(3)}
=
\left[
\begin{array}{cc}
1&\\
&2I_7\\
\end{array}
\right].
\end{align}
Thus, we have seven $\Z_2$-valued SIs $\nu^{(3)}_2,\dots,\nu^{(3)}_8$ detecting 3nd-order TSCs. 

The SIs can be written as~\footnote{Here we took unimodular transformations to make the formulas of the SIs simple.}
\begin{align}
\begin{array}{cc|cccccccc|l}
{\rm SI}&{\rm Range}&n^{000}&n^{100}&n^{010}&n^{001}&n^{110}&n^{011}&n^{101}&n^{111}&\mbox{Target}\\
\hline
\nu^{(3)}_2&\Z/2\Z& 1 & -1 & 0 & 0 & 0 & 0 & 0 & 0 &\mbox{3rd-order TSC}\\
\nu^{(3)}_3&\Z/2\Z& 1 & 0 & -1 & 0 & 0 & 0 & 0 & 0 \\
\nu^{(3)}_4&\Z/2\Z& 1 & 0 & 0 & -1 & 0 & 0 & 0 & 0 \\
\nu^{(3)}_5&\Z/2\Z& \frac{1}{2} & -\frac{1}{2} & -\frac{1}{2} & 0 & \frac{1}{2} & 0 & 0 & 0 &\\
\nu^{(3)}_6&\Z/2\Z& \frac{1}{2} & 0 & -\frac{1}{2} & -\frac{1}{2} & 0 & \frac{1}{2} & 0 & 0 \\
\nu^{(3)}_7&\Z/2\Z& \frac{1}{2} & -\frac{1}{2} & 0 & -\frac{1}{2} & 0 & 0 & \frac{1}{2} & 0 \\
\nu^{(3)}_8&\Z/2\Z& \frac{1}{4} & -\frac{1}{4} & -\frac{1}{4} & -\frac{1}{4} & \frac{1}{4} & \frac{1}{4} & \frac{1}{4} & -\frac{1}{4} &\\
\hline
\nu^{(2)}_5&\Z/2\Z& 1 & -1 & -1 & 0 & 1 & 0 & 0 & 0 &\mbox{2nd-order TSC}\\
\nu^{(2)}_6&\Z/2\Z& 1 & 0 & -1 & -1 & 0 & 1 & 0 & 0 \\
\nu^{(2)}_7&\Z/2\Z& 1 & -1 & 0 & -1 & 0 & 0 & 1 & 0 \\
\nu^{(2)}_8&\Z/2\Z& \frac{1}{2} & -\frac{1}{2} & -\frac{1}{2} & -\frac{1}{2} & \frac{1}{2} & \frac{1}{2} & \frac{1}{2} & -\frac{1}{2} &\\
\hline
\nu^{(1)}_8 &\Z/2\Z& 1 & -1 & -1 & -1 & 1 & 1 & 1 & -1 &\mbox{1st-order TSC} \\
\end{array}
\end{align}
Note that the SI $\nu^{(1)}_8=2\nu^{(2)}_8=4\nu^{(3)}_8$ eventually becomes a $\Z/8\Z$-valued SI. 
Similally, the SIs $\nu^{(2)}_i=2\nu^{(3)}_i \ (i=5,6,7,8)$ become $\Z/\Z_4$-valued SIs. 

It should be noted that the same SIs are obtained by the quotient $E_1^{0,0}/\im f^{\rm TI}_{d=0}$. 
The SNF of $M_{f^{\rm TI}_{d=0}}$ is given as 
\begin{align}
u M_{f^{\rm TI}_{d=0}} v=
\left[
\begin{array}{cccc}
1\\
&2I_3\\
&&4I_3\\
&&&8\\
\end{array}
\right].
\end{align}
We have the SIs $\nu_i = \sum_{j=1}^8 u_{ij} n_j$ with 
\begin{align}
u=
\begin{array}{cc|cccccccc}
\mbox{SI}&\mbox{Range}&n^{000}&n^{100}&n^{010}&n^{001}&n^{110}&n^{011}&n^{101}&n^{111}\\
\hline
\nu_1&\Z/\Z&1 & 0 & 0 & 0 & 0 & 0 & 0 & 0 \\
\hline
\nu_2&\Z/2\Z& 1 & -1 & 0 & 0 & 0 & 0 & 0 & 0 \\
\nu_3&\Z/2\Z& 1 & 0 & -1 & 0 & 0 & 0 & 0 & 0 \\
\nu_4&\Z/2\Z& 1 & 0 & 0 & -1 & 0 & 0 & 0 & 0 \\
\hline
\nu_5&\Z/4\Z& 1 & -1 & -1 & 0 & 1 & 0 & 0 & 0 \\
\nu_6&\Z/4\Z& 1 & 0 & -1 & -1 & 0 & 1 & 0 & 0 \\
\nu_7&\Z/4\Z& 1 & -1 & 0 & -1 & 0 & 0 & 1 & 0 \\
\hline
\nu_8&\Z/8\Z& 1 & -1 & -1 & -1 & 1 & 1 & 1 & -1 \\
\end{array}. 
\end{align}
From the expression (\ref{eq:3dtsc_top_inv_wcl}) in the weak coupling limit, the SIs can be written only with the normal state. 
For instance, 
\begin{equation}\begin{split}
\nu_8
&=
\frac{1}{2}
\Big[
\big\{
N_+[h(0,0,0)]
-N_+[h(\pi,0,0)]
-N_+[h(0,\pi,0)]
-N_+[h(0,0,\pi)]\\
&\qquad \quad +N_+[h(\pi,\pi,0)]
+N_+[h(\pi,0,\pi)]
+N_+[h(0,\pi,\pi)]
-N_+[h(\pi,\pi,\pi)]
\big\}\\
&\quad \quad -
\big\{
N_-[h(0,0,0)]
-N_-[h(\pi,0,0)]
-N_-[h(0,\pi,0)]
-N_-[h(0,0,\pi)]\\
&\qquad \quad +N_-[h(\pi,\pi,0)]
+N_-[h(\pi,0,\pi)]
+N_-[h(0,\pi,\pi)]
-N_-[h(\pi,\pi,\pi)]
\big\}
\Big]. 
\end{split}\end{equation}
The essentially same SIs $\nu_i$s were derived in Ref.~\cite{SkurativskaSISC}. 

In the same way as in Secs.~\ref{sec:2d_c4_ai1}, \ref{sec:2d_c4_ai0}, taking the quotient by atomic insulators localized at the interior of the unit cell, one can define the SIs for Andreev bound states. 
We do not repeat here.

\section{Summary}
In this note, we depicted two routes for generalizing the SI for electric material in Refs.[Po,Haruki]. 
The one is the SIs for superconductors, another one is the SIs for ingap boundary states. 
To do so, in Sec.~\ref{sec:formulation}, we first empathized that there exists a filtration (\ref{eq:subgroup_str}) of the group $E_1^{0,0}$ of topological invariants at high-symmetric points, which is originated from various definitions of nontrivial topology such as semimetal phases, higher-order TIs/TSCs and ingap boundary states. 
We illustrated how the explicit formulas of the SIs are constructed in Sec.~\ref{sec:formulation}, in the cases where the group $E_1^{0,0}$ is free abelian.
For SCs, there emerges a new family of SIs beyond those for electric materials, from the careful definition of what atomic insulators and the trivial vacuum Hamiltonian are. 
We also showed that taking the quotient of the group $E_1^{0,0}$ by the subset of atomic insulators such as ones localized at the interior of the unit cell yields the SIs for detecting ingap corner, hinge, and boundary states. 
We demonstrated our framework does work for a few examples in in Sec.~\ref{sec:exs}. 
We leave the comprehensive classification of the SIs for SCs and ingap boundary states as a future work.

\noindent
{\it Acknowledgement---}
We thank 
Akira Furusaki, 
Motoaki Hirayama, 
Tomoki Ozawa, 
Masatoshi Sato, 
Luka Trifunovic, 
Youichi Yanase, 
and 
Tiantian Zhang 
for helpful discussions.
We especially thank Yasuhiro Ishitsuka for teaching me how to compute $\coker f, \ker f, \im f$ of a homomorphism $f$ of abelian groups. 
This work was supported by PRESTO, JST (Grant No. JPMJPR18L4).

\appendix

\section{On the computation of a homomorphism $f:A \to B$}
\label{app:homo}
Let $A, B$ finitely generated abelian groups, and $f: A\to B$ be a homomorphism between them. 
We shall compute $\ker f, \im f$, and ${\rm Coker} f$. 
The strategy is as follows.~\cite{Yasuhiro Ishitsuka, private communication.}
Given an arbitrary abelian group $A$, there exists a free abelian group $F$ and a surjective group homomorphism $\pi_A: F \to A$. 
The homomorphism $f: A \to B$ can be lifted to a homomorphism $\tilde f: F \to G$ between free abelian groups. 
The Smith normal form of the representation matrix of $\tilde f$ and the inclusions $i_A: \ker \pi_A \to F$, $i_B: \ker \pi_B \to G$ give us the desired groups. 

Let us denote the bases of $A, B$ by $\{a_j\}_{j=1}^n, \{b_j\}_{j=1}^M$, respectevely, and $F=\bigoplus_{j=1}^n \Z[\tilde a_j]$, $G=\bigoplus_{j=1}^m \Z[\tilde b_j]$ for the integral lifts. 
We have the commutative diagram 
\begin{align}
\begin{CD}
@. 0@.0\\
@. @VVV @VVV \\
\ker \tilde f|_{{\rm ker} \pi_A}@>>> \ker \pi_A @>\tilde f|_{{\rm ker} \pi_A}>> \ker \pi_B @>>> {\rm Coker} \tilde f|_{{\rm ker} \pi_A}\\
@VVV @Vi_AVV @Vi_BVV @VVV \\
\ker \tilde f @>>> \bigoplus_j \Z[\tilde a_j] @>\tilde f>> \bigoplus_j \Z[\tilde b_j] @>>> {\rm Coker}\tilde f \\
@VVV @V\pi_AVV @V\pi_BVV @VVV\\
\ker f@>>> A @>f>> B @>>> {\rm Coker} f \\
@. @VVV @VVV \\
@. 0@.0\\
\end{CD}.
\label{eq:comm_diagram}
\end{align}
The homomorphism $\tilde f$ and inclusions $i_A,i_B$ are given as follows. 
Let us write 
\begin{align}
&A = \bigoplus_{j=1}^k \Z[a_j] \oplus \bigoplus_{j=k+1}^n \Z_{p_j}[a_j], \qquad 
\tilde A = \bigoplus_{j=1}^k \Z[\tilde a_j] \oplus \bigoplus_{j=k+1}^n \Z[\tilde a_j], \\
&B = \bigoplus_{j=1}^l \Z[b_j] \oplus \bigoplus_{j=l+1}^m \Z_{q_j}[b_j], \qquad 
\tilde B = \bigoplus_{j=1}^l \Z[\tilde b_j] \oplus \bigoplus_{j=l+1}^m \Z[\tilde b_j]. 
\end{align}
The inclusions $i_A,i_B$ are given as 
\begin{align}
&\ker \pi_A = \bigoplus_{j=k+1}^n \Z[\tilde a_j], \nonumber \\
&i_A(\tilde a_{k+1},\dots,\tilde a_{n})
= (\tilde a_1,\dots,\tilde a_k;\tilde a_{k+1},\dots,\tilde a_n) M_{i_A}, \qquad 
M_{i_A}=
\left[
\begin{array}{cccc}
&O&\\
\hline
p_{k+1}&&\\
&\ddots \\
&&p_n\\
\end{array}
\right], \\
&\ker \pi_B = \bigoplus_{j=l+1}^m \Z[\tilde b_j], \nonumber \\
&i_B(\tilde b_{l+1},\dots,\tilde b_{m})
= (\tilde b_1,\dots,\tilde b_l;\tilde b_{l+1},\dots,\tilde b_m)M_{i_B}, \qquad 
M_{i_B}=
\left[
\begin{array}{ccc}
&O&\\
\hline
q_{l+1}&&\\
&\ddots \\
&&q_m\\
\end{array}
\right].
\end{align}
The homomorphism $f$ is represented as 
\begin{align}
f(a_1,\dots, a_k; a_{k+1},\dots, a_n)
=
(b_1,\dots, b_l; b_{l+1},\dots, b_m) 
\left[
\begin{array}{c|ccc}
M && O & \\
\hline
\alpha_{l+1} & \beta_{k+1,l+1} & \cdots & \beta_{k+1,n} \\
\vdots&\vdots & & \vdots \\
\alpha_{m} & \beta_{m,l+1} & \cdots & \beta_{m,n} \\
\end{array}
\right], 
\end{align}
where 
\begin{align}
&M \in {\rm Mat}_{l\times k}(\Z), \nonumber\\
&\alpha_j \in {\rm Mat}_{1 \times k}(\Z_{q_j}), \qquad  (j=l+1,\dots m), \nonumber\\
&\beta_{ij} \in {\rm Hom}(\Z_{p_i},\Z_{q_j}) = (q/{\rm gcd}_{p_i,q_j})\Z/q_j\Z (\cong \Z_{{\rm gcd}_{p_i,q_j}}), \qquad (i=k+1,\dots,m, j=l+1,\dots,n).
\end{align}
The homomorphism $\tilde f$ is given by integral lifts of matrix elements 
\begin{align}
&\tilde f(\tilde a_1,\dots, \tilde a_k; \tilde a_{k+1},\dots, \tilde a_n)
=
(\tilde b_1,\dots, \tilde b_l; \tilde b_{l+1},\dots, \tilde b_m) M_{\tilde f}, \nonumber \\
&M_{\tilde f}=
\left[
\begin{array}{c|ccc}
M && O & \\
\hline
\tilde \alpha_{l+1} & \tilde \beta_{k+1,l+1} & \cdots & \tilde \beta_{k+1,n} \\
\vdots&\vdots & & \vdots \\
\tilde \alpha_{m} & \tilde \beta_{m,l+1} & \cdots & \tilde \beta_{m,n} \\
\end{array}
\right], 
\end{align}
where 
\begin{align}
&\alpha_j \mapsto \tilde \alpha_j \in {\rm Mat}_{1 \times k}(\Z), \qquad  (j=l+1,\dots m), \nonumber \\
&\beta_{ij} \mapsto \tilde \beta_{ij} \in\Z, \qquad (i=k+1,\dots,m, j=l+1,\dots,n).
\end{align}

\subsection{${\rm Coker} f$}
The quotient group ${\rm Coker} f = B/\im f$ is given by the quotient group $\bigoplus_j \Z[\tilde b_j] / \im \tilde f$ modulo $i_B (\ker \pi_B)$.
Therefore, ${\rm Coker} f$ is written as~\cite{Ishitsuka}
\begin{align}
{\rm Coker} f
=
\bigoplus_j \Z[\tilde b_j] \Big/\big(\im \tilde f + i_B(\ker \pi_B)\big), 
\end{align}
where $X+Y$ is $X \cup Y$ us a set. 
To compute $\big(\im \tilde f + i_B(\ker \pi_B)\big)$, introduce the homomorphism 
\begin{align}
&\tilde f \oplus i_B: \tilde A \oplus \ker B \to \tilde B, \nonumber \\
&(\tilde f \oplus i_B)(\tilde a_1,\dots, \tilde a_n;\tilde b_{l+1},\dots,\tilde b_m)
=
(\tilde b_1,\dots,\tilde b_l;\tilde b_{l+1},\dots,\tilde b_m) M_{\tilde f \oplus i_B}, \nonumber \\
&M_{\tilde f \oplus i_B}
=
\left[
\begin{array}{c|ccc|ccccc}
M && O &&&O\\
\hline
\tilde \alpha_{l+1} & \tilde \beta_{k+1,l+1} & \cdots & \tilde \beta_{k+1,n}&q_{l+1}&&\\
\vdots&\vdots & & \vdots &&\ddots &\\
\tilde \alpha_{m} & \tilde \beta_{m,l+1} & \cdots & \tilde \beta_{m,n}&&&q_m \\
\end{array}
\right].
\end{align}
Applying the Smith decomposition to the matrix $M_{\tilde f \oplus i_B}$, we have 
\begin{align}
&u M_{\tilde f \oplus i_B} v = 
\left[
\begin{array}{cc}
D_\lambda&O\\
O&O
\end{array}
\right], \qquad 
D_\lambda=
\left[
\begin{array}{ccc}
\lambda_1&&\\
&\ddots&\\
&&\lambda_r\\
\end{array}
\right], 
\label{eq:SNF_tildef+iB}
\end{align}
\begin{align}
(\tilde f \oplus i_B)(\tilde a_1,\dots, \tilde a_n;\tilde b_{l+1},\dots,\tilde b_m)v
=
(\tilde b_1,\dots,\tilde b_l;\tilde b_{l+1},\dots,\tilde b_m)u^{-1}
\left[
\begin{array}{ccc|c}
\lambda_1&&&\\
&\ddots&&O\\
&&\lambda_r&\\
\hline 
&O&&O\\
\end{array}
\right],
\end{align}
where $\lambda_i (i=1,\dots r)$ are nonnegative integers, and $u,v$ are unimodular matrices. 
Introducing the new basis of $\bigoplus_j \Z[\tilde b_j]$ by 
\begin{align}
(\tilde b'_1,\dots, \tilde b'_m) = 
(\tilde b_1,\dots, \tilde b_m) u^{-1}, 
\label{eq:app_btildeprime}
\end{align}
we see that 
\begin{align}
\im \tilde f + i_B(\ker \pi_B)
= 
\bigoplus_{j=1}^r \Z[\lambda_j \tilde b'_j], 
\end{align}
and 
\begin{align}
\coker f \cong 
\bigoplus_{j=1}^r \Z_{\lambda_j}[\tilde b'_j] \oplus \bigoplus_{j=r+1}^m \Z[\tilde b'_j]. 
\end{align}

\subsection{$\ker f$}
The group $\ker f$ can be computed as~\cite{Ishitsuka}
\begin{align}
\ker f 
=
\tilde f^{-1}(i_B \ker \pi_B)/i_A \ker \pi_A.
\end{align}
Here note that, from the commutative diagram (\ref{eq:comm_diagram}), for $i_A(x) \in i_A \ker \pi_A$ we have $\tilde f(x) = i_B (\tilde f|_{\ker \pi_A}(x)) \in i_B(\ker \pi_B)$, thus $i_A \ker \pi_A \subset \tilde f^{-1}(i_B \ker \pi_B)$. 

Let us first compute $\tilde f^{-1}(i_B \ker \pi_B)$. 
By using the homomorphism $\tilde f \oplus i_B$ introduced before, we have 
\begin{align}
\tilde f^{-1}(i_B \ker \pi_B)
=
\pi_{\tilde A} \Big( \ker (\tilde f \oplus i_B )\Big), 
\end{align}
where $\pi_{\tilde A}: \tilde A \oplus \ker B \to \tilde A$ is the projection. 
Using the Smith normal form (\ref{eq:SNF_tildef+iB}), $\ker (\tilde f \oplus i_B )$ is spanned by 
\begin{align}
\sum_{i=1}^n \tilde a_i v_{ij} + \sum_{i=1}^{m-l} \tilde b_i v_{n+i,j}, \qquad j=r+1,\dots,n+m-l. 
\end{align}
Therefore, $\pi_{\tilde A} \Big( \ker (\tilde f \oplus i_B )\Big)$ is generated by elements 
\begin{align}
\sum_{i=1}^n \tilde a_i v_{ij}, \qquad j=r+1,\dots,n+m-l. 
\end{align}
Let us write 
\begin{align}
v_{\rm sub} = 
\left[ 
\begin{array}{cccccc}
v_{1,r+1}&\cdots&v_{1,n+m-l}\\
\vdots&&\vdots\\
v_{n,r+1}&\cdots&v_{n,n+m-l}\\
\end{array}
\right].
\end{align}
Applying the Smith decomposition to $v_{\rm sub}$, we have 
\begin{align}
u' v_{\rm sub} v' = 
\left[
\begin{array}{c|c}
D'&O\\
\hline
O&O\\
\end{array}
\right], \qquad 
D'=\left[
\begin{array}{ccc}
d'_1&&\\
&\ddots&\\
&&d'_s\\
\end{array}
\right]. 
\end{align}
In the new basis 
\begin{align}
(\tilde a'_1, \dots, \tilde a'_n):=(\tilde a_1,\dots,\tilde a_n) u'^{-1} 
\end{align}
of $\tilde A$, we have 
\begin{align}
\pi_{\tilde A} \Big( \ker (\tilde f \oplus i_B )\Big)
=
\bigoplus_{j=1}^s \Z[d'_j \tilde a_j']. 
\end{align}

The quotient 
\begin{align}
\ker f
= 
\bigoplus_{j=1}^s \Z[d'_j \tilde a_j'] \Big/ i_A \ker \pi_A 
\end{align}
is given as follows. 
We examine the inclusion $i_A$ in the basis of $\{\tilde a_j'\}_{j=1}^n$, 
\begin{align}
i_A(\tilde a_{k+1},\dots, \tilde a_n)
=(\tilde a_1,\dots,\tilde a_n) M_{i_A}
=(\tilde a_1',\dots,\tilde a_n') u' M_{i_A}. 
\end{align}
Since $i_A \ker \pi_A \subset \bigoplus_{j=1}^s \Z[d'_j \tilde a_j']$, the r.h.s.\ is written as 
\begin{align}
&i_A(\tilde a_{k+1},\dots, \tilde a_n)
=(\tilde a_1,\dots,\tilde a_n) M_{i_A}
=(d'_1\tilde a_1',\dots,d'_s\tilde a_s';\tilde a'_{s+1},\dots,\tilde a'_n) 
\left[
\begin{array}{c}
M'\\
O
\end{array}
\right], \nonumber\\
&M' \in {\rm Mat}_{s \times (n-k)}(\Z), \qquad 
\left[
\begin{array}{c}
M'\\
O
\end{array}
\right]
=
\left[
\begin{array}{cc}
D'^{-1} & O \\
O & I_{n-s} \\
\end{array}
\right]
u'M_{i_A}.
\end{align}
The final step is to apply the Smith decomposition to $M'$. 
We have 
\begin{align}
u^{(1)} M' v^{(1)} = 
\left[
\begin{array}{ccc|c}
d_1^{(1)}&&&\\
&\ddots&&O\\
&&d^{(1)}_{s_1}&\\
\hline
&O&&O
\end{array}
\right], 
\end{align}
\begin{align}
&i_A(\tilde a_{k+1},\dots, \tilde a_n) v^{(1)}
=(d'_1\tilde a_1',\dots,d'_s\tilde a_s') [u^{(1)}]^{-1} 
\left[
\begin{array}{ccc|c}
d_1^{(1)}&&&\\
&\ddots&&O\\
&&d^{(1)}_{s_1}&\\
\hline
&O&&O
\end{array}
\right].
\end{align}
Let us introduce the basis 
\begin{align}
(\tilde a^{(1)}_1,\dots,\tilde a^{(1)}_s)=(d'_1\tilde a_1',\dots,d'_s\tilde a_s') [u^{(1)}]^{-1}. 
\end{align}
Then, we arrive at 
\begin{align}
\ker f
=
\bigoplus_{j=1}^{s_1} \Z_{d^{(1)}_j}[\tilde a^{(1)}_j] \oplus \bigoplus_{j=s_1+1}^s \Z[\tilde a^{(1)}_j]. 
\end{align}

%

\subsection{$\im f$}
The image of $f$ is written as~\cite{Ishitsuka}
\begin{align}
\im f 
= 
\im \tilde f/ \left( \im \tilde f \cap i_B(\ker \pi_B) \right)
=
\left( \im \tilde f + i_B(\ker \pi_B) \right)/i_B(\ker \pi_B).
\end{align}
The inclusion $i_B(\ker \pi_B) \to \im \tilde f + i_B(\ker \pi_B)$ is computed as follows. 
In the basis $\{\tilde b'_j\}_{j=1}^m$ introduced in (\ref{eq:app_btildeprime}), $i_B$ is written as 
\begin{align}
i_B (\tilde b_{l+1},\dots,\tilde b_m)
&=
(\tilde b'_1,\dots,\tilde b'_m) u M_{i_B} \nonumber\\
&=
(\lambda_1\tilde b'_1,\dots,\lambda_r\tilde b'_r;\tilde b'_{r+1},\dots,\tilde b'_m) 
\left[
\begin{array}{cc}
D_\lambda^{-1}&O\\
O&I_{m-r}\\
\end{array}
\right]
u M_{i_B} \nonumber\\
&=
(\lambda_1\tilde b'_1,\dots,\lambda_r\tilde b'_r;\tilde b'_{r+1},\dots,\tilde b'_m) 
\left[
\begin{array}{c}
M''\\
O\\
\end{array}
\right], 
\end{align}
where $M'' \in {\rm Mat}_{r \times (m-l)}(\Z)$.  
Consider the Smith normal form of $M''$, 
\begin{align}
u^{(2)} M'' v^{(2)} = 
\left[
\begin{array}{ccc|c}
d^{(2)}_1&&&\\
&\ddots&&O\\
&&d^{(2)}_{r_1}&\\
\hline 
&O&&O\\
\end{array}
\right]. 
\end{align}
We have 
\begin{align}
\Big( \im \tilde f + i_B(\ker \pi_B) \Big)/i_B \ker \pi_B
= 
\bigoplus_{j=1}^{r_1} \Z_{d^{(2)}_j}[\lambda_j \tilde b'_j] \oplus 
\bigoplus_{j=r_1+1}^r \Z[\lambda_j \tilde b'_j].
\end{align}
%
%
%
%
%
%


\bibliography{refs}

\begin{thebibliography}{10}

\bibitem{Po230}
Hoi~Chun Po, Ashvin Vishwanath, and Haruki Watanabe.
\newblock Complete theory of symmetry-based indicators of band topology.
\newblock {\em Nat. Commun.}, 8(1):50, 2017.

\bibitem{Haruki1651}
Haruki Watanabe, Hoi~Chun Po, and Ashvin Vishwanath.
\newblock Structure and topology of band structures in the 1651 magnetic space
  groups.
\newblock {\em Science Advances}, 4(8), 2018.

\bibitem{FuKaneInversion}
Liang Fu and C.~L. Kane.
\newblock Topological insulators with inversion symmetry.
\newblock {\em Phys. Rev. B}, 76:045302, Jul 2007.

\bibitem{FuBerg}
Liang Fu and Erez Berg.
\newblock Odd-parity topological superconductors: Theory and application to
  ${\mathrm{cu}}_{x}{\mathrm{bi}}_{2}{\mathrm{se}}_{3}$.
\newblock {\em Phys. Rev. Lett.}, 105:097001, Aug 2010.

\bibitem{SatoOddParity}
Masatoshi Sato.
\newblock Topological odd-parity superconductors.
\newblock {\em Phys. Rev. B}, 81:220504, Jun 2010.

\bibitem{ChenRotation}
Chen Fang, Matthew~J. Gilbert, and B.~Andrei Bernevig.
\newblock Bulk topological invariants in noninteracting point group symmetric
  insulators.
\newblock {\em Phys. Rev. B}, 86:115112, Sep 2012.

\bibitem{KS_Atiyah}
Ken Shiozaki, Masatoshi Sato, and Kiyonori Gomi.
\newblock Atiyah-hirzebruch spectral sequence in band topology: General
  formalism and topological invariants for 230 space groups.
\newblock arXiv:1802.06694.

\bibitem{ChenGlide}
Chen Fang and Liang Fu.
\newblock New classes of three-dimensional topological crystalline insulators:
  Nonsymmorphic and magnetic.
\newblock {\em Phys. Rev. B}, 91:161105, Apr 2015.

\bibitem{KSGlide}
Ken Shiozaki, Masatoshi Sato, and Kiyonori Gomi.
\newblock ${Z}_{2}$ topology in nonsymmorphic crystalline insulators: M\"obius
  twist in surface states.
\newblock {\em Phys. Rev. B}, 91:155120, Apr 2015.

\bibitem{KSGlide2}
Ken Shiozaki, Masatoshi Sato, and Kiyonori Gomi.
\newblock Topology of nonsymmorphic crystalline insulators and superconductors.
\newblock {\em Phys. Rev. B}, 93:195413, May 2016.

\bibitem{RyuTenFold}
Shinsei Ryu, Andreas~P. Schnyder, Akira Furusaki, and Andreas~W.W. Ludwig.
\newblock Topological insulators and superconductors: tenfold way and
  dimensional hierarchy.
\newblock {\em New Journal of Physics}, 12(6):065010, 2010.

\bibitem{KitaevPeriodic}
Alexei Kitaev.
\newblock Periodic table for topological insulators and superconductors.
\newblock {\em AIP Conference Proceedings}, 1134(1):22--30, 2009.

\bibitem{TQC}
Barry Bradlyn, L~Elcoro, Jennifer Cano, MG~Vergniory, Zhijun Wang, C~Felser,
  MI~Aroyo, and B~Andrei Bernevig.
\newblock Topological quantum chemistry.
\newblock {\em Nature}, 547(7663):298, 2017.

\bibitem{SongAII}
Zhida Song, Tiantian Zhang, Zhong Fang, and Chen Fang.
\newblock Quantitative mappings between symmetry and topology in solids.
\newblock {\em Nature communications}, 9(1):3530, 2018.

\bibitem{SongAI}
Zhida Song, Tiantian Zhang, and Chen Fang.
\newblock Diagnosis for nonmagnetic topological semimetals in the absence of
  spin-orbital coupling.
\newblock {\em Phys. Rev. X}, 8:031069, Sep 2018.

\bibitem{KhalafPRX18AII}
Eslam Khalaf, Hoi~Chun Po, Ashvin Vishwanath, and Haruki Watanabe.
\newblock Symmetry indicators and anomalous surface states of topological
  crystalline insulators.
\newblock {\em Phys. Rev. X}, 8:031070, Sep 2018.

\bibitem{OnoClassA}
Seishiro Ono and Haruki Watanabe.
\newblock Unified understanding of symmetry indicators for all internal
  symmetry classes.
\newblock {\em Phys. Rev. B}, 98:115150, Sep 2018.

\bibitem{ZhangCatalogue}
Tiantian Zhang, Yi~Jiang, Zhida Song, He~Huang, Yuqing He, Zhong Fang, Hongming
  Weng, and Chen Fang.
\newblock Catalogue of topological electronic materials.
\newblock {\em Nature}, 566(7745):475, 2019.

\bibitem{VergnioryCatalogue}
MG~Vergniory, L~Elcoro, Claudia Felser, Nicolas Regnault, B~Andrei Bernevig,
  and Zhijun Wang.
\newblock A complete catalogue of high-quality topological materials.
\newblock {\em Nature}, 566(7745):480, 2019.

\bibitem{TangCatalogue}
Feng Tang, Hoi~Chun Po, Ashvin Vishwanath, and Xiangang Wan.
\newblock Comprehensive search for topological materials using symmetry
  indicators.
\newblock {\em Nature}, 566(7745):486, 2019.

\bibitem{OnoSIforSC}
Seishiro Ono, Youichi Yanase, and Haruki Watanabe.
\newblock Symmetry indicators for topological superconductors.
\newblock arXiv:1811.08712.

\bibitem{SkurativskaSISC}
Anastasiia Skurativska, Titus Neupert, and Mark~H Fischer.
\newblock Atomic limit and inversion-symmetry indicators for topological
  superconductors.
\newblock arXiv:1906.11267.

\bibitem{BenalcazarCorner}
Wladimir~A. Benalcazar, Tianhe Li, and Taylor~L. Hughes.
\newblock Quantization of fractional corner charge in ${C}_{n}$-symmetric
  higher-order topological crystalline insulators.
\newblock {\em Phys. Rev. B}, 99:245151, Jun 2019.

\bibitem{SchindlerCorner}
Frank Schindler, Marta Brzezi{\'n}ska, Wladimir~A Benalcazar, Mikel Iraola,
  Adrien Bouhon, Stepan~S Tsirkin, Maia~G Vergniory, and Titus Neupert.
\newblock Fractional corner charges in spin-orbit coupled crystals.
\newblock arXiv:1907.10607.

\bibitem{PhysRevX.7.041069}
Jorrit Kruthoff, Jan de~Boer, Jasper van Wezel, Charles~L. Kane, and Robert-Jan
  Slager.
\newblock Topological classification of crystalline insulators through band
  structure combinatorics.
\newblock {\em Phys. Rev. X}, 7:041069, Dec 2017.

\bibitem{BenalcazarScience}
Wladimir~A. Benalcazar, B.~Andrei Bernevig, and Taylor~L. Hughes.
\newblock Quantized electric multipole insulators.
\newblock {\em Science}, 357(6346):61--66, 2017.

\bibitem{karoubi2008k}
Max Karoubi.
\newblock {\em K-theory: An introduction}, volume 226.
\newblock Springer Science \& Business Media, 2008.

\bibitem{kitaev2001unpaired}
A~Yu Kitaev.
\newblock Unpaired majorana fermions in quantum wires.
\newblock {\em Physics-Uspekhi}, 44(10S):131, 2001.

\bibitem{ono2019difficulties}
Seishiro Ono, Luka Trifunovic, and Haruki Watanabe.
\newblock Difficulties in operator-based formulation of the bulk quadrupole
  moment.
\newblock {\em arXiv preprint arXiv:1902.07508}, 2019.

\end{thebibliography}
\bibliographystyle{unsrt} 

\end{document}